\documentclass{article}%
\usepackage{pbox}
\usepackage{graphicx}
\usepackage{amsmath}
\usepackage{amssymb}
\usepackage{subfigure}
\usepackage{verbatim}
\usepackage{epstopdf}
\usepackage[affil-it]{authblk}
\usepackage{float,color}
\usepackage{amsfonts}
\usepackage[margin=1in]{geometry}%
\usepackage[title]{appendix}
\providecommand{\U}[1]{\protect\rule{.1in}{.1in}}
\newtheorem{exmp}{Example}[section]

\renewcommand\thesection{\arabic{section}}

\renewcommand\theequation{\thesection.\arabic{equation}}
\renewcommand{\eqref}[1]{(\ref{#1})}
\usepackage[numbers,sort&compress]{natbib}

\newcommand{\eps}{{\displaystyle \varepsilon}}
\newcommand{\bsub}{\begin{subequations}}
\newcommand{\esub}{\end{subequations}$\!$}
\newcommand{\BE}{\begin{equation}}
\newcommand{\EE}{\end{equation}}

\newcommand{\bs}[0]{\boldsymbol}

\newcommand{\bigoh}{\mathcal{O}}

\newcommand{\bx}{\mathbf{x}}

\newcommand{\br}{\mathbf{r}}
\newcommand{\by}{\mathbf{y}}
\newcommand{\bn}{\mathbf{n}}

\newcommand{\bA}{\mathbf{A}}
\newcommand{\bg}{\mathbf{g}}
\newcommand{\bxi}{\bs \xi}
\newcommand{\noI}{\noindent}

\newcommand{\al}[1]{\textcolor{black}{#1}}
\newcommand{\jt}[1]{\textcolor{black}{#1}}

\begin{document}

\title{Hybrid asymptotic-numerical approach for estimating first passage time densities of the two-dimensional narrow capture problem}
\author{A.~E.~Lindsay\thanks{Department of Applied \& Computational Mathematics \& Statistics, University of Notre Dame, Notre Dame, IN, 46556, USA, {\tt a.lindsay@nd.edu}}\enspace , R.~T.~Spoonmore\thanks{Department of Applied \& Computational Mathematics \& Statistics, University of Notre Dame, Notre Dame, IN, 46556, USA, {\tt Ryan.T.Spoonmore.2@nd.edu}} \enspace J.~C.~Tzou\thanks{Mathematics Department, University of British Columbia, Vancouver, BC, Canada {\tt tzou.justin@gmail.com}} }
\maketitle

\begin{abstract}
A hybrid asymptotic-numerical method is presented for obtaining \jt{an asymptotic estimate for} the full probability distribution of capture times of a random walker by multiple small traps located inside a bounded two-dimensional domain with a reflecting boundary. As motivation for this study, we calculate the variance in the capture time of a random walker by a single interior trap and determine this quantity to be comparable in magnitude to the mean. This implies that the mean is not necessarily reflective of typical capture times and that the full density must be determined. To solve the underlying diffusion equation, the method of Laplace transforms is used to obtain an elliptic problem of modified Helmholtz type. In the limit of vanishing trap sizes, each trap is represented as a Dirac point source which permits the solution of the transform equation to be represented as a superposition of Helmholtz Green's functions. Using this solution, we construct asymptotic short time solutions of the first passage time density which captures peaks associated with rapid capture by the absorbing traps. When numerical evaluation of the Helmholtz Green's function is employed followed by numerical inversion of the Laplace transform, the method reproduces the density for larger times. We demonstrate the accuracy of our solution technique with comparison to statistics obtained from a time-dependent solution of the diffusion equation and discrete particle simulations. In particular, we demonstrate that the method is capable of capturing the multimodal behavior in the capture time density that arises when the traps are strategically arranged. The hybrid method presented can be applied to scenarios involving both arbitrary domains and trap shapes.
\end{abstract}


\label{firstpage}

\setcounter{equation}{0}
\section{Introduction}

\al{In many biological, social and physical processes, the arrival of a single individual or particle at a reaction site can initiate a cascade of events. When the dynamics of these particles are driven by random motions, the distribution of arrival times is known as the first passage time density. This distribution and its moments give crucial information on the feasibility, effectiveness and robustness of stochastic transport mechanisms \cite{Newby2013}. In biologically motivated first passage time problems, the number of individual particles can be very large and the target site(s) relatively small compared to the total search domain.} \jt{The so-called narrow escape problem in two and three dimensions seeks the average time required for such a particle to reach a target site and has been the subject of intensive study (see e.g., \cite{HolcmanReview2014, holcman2015stochastic} for a comprehensive review of the associated techniques and applications). For example, the rate of escape of ions through ion channels located on a cell membrane may yield insights into the relevant timescales over which cellular processes occur \cite{Holcman2007PNAS}. In three dimensions, the amount of time it takes for a T cell to find its antigen may be indicative of immune response times \cite{delgado2015conditional}. In some applications, however, the mean of the trapping time may not yield sufficient information. In a random walk model of conformational transitions in proteins, it was shown under certain scenarios that the probability of the protein residing within a certain class of states is highly dependent on its initial state and the subsequent early time evolution \cite{kurzynski1998time}. While the first moment of the passage time density yields information regarding the tail of the first passage time distribution (large times), it neglects  information on the small time distribution required to fully characterize the transitional dynamics \cite{kurzynski1998time}.}

\jt{In this work, w}e present a methodology for determining the full distribution of first passage times of a diffusing particle in a two dimensional bounded spatial region $\Omega$ to reach a small absorbing set $\Omega_{\eps}$.  While diffusion accompanied by directed motion is often important in such problems, for simplicity of exposition we consider unbiased Brownian motion. The probability $p(\bx,t; \bxi)$ that the particle initially at $\bxi\in\Omega$ is free at $\bx \in\Omega$ at time $t>0$ satisfies the diffusion equation
\bsub\label{MainFreeProb}
\begin{gather}
\label{MainFreeProb_a} \frac{\partial p}{\partial t} = \Delta p, \qquad t>0, \quad \bx\in\Omega\setminus\Omega_{\eps}; \qquad \partial_n p = 0, \qquad t>0 \quad \bx \in \partial \Omega;\\[5pt]
\label{MainFreeProb_b} p = 0, \qquad t>0, \quad \bx \in \partial\Omega_{\eps}, \qquad p(\bx,0;\bxi) = \delta(\bx - \bxi ), \quad \bx \in \Omega \,,
\end{gather}
\esub
in a bounded region $\Omega\subset\mathbb{R}^d$, $d = 1,2,3$ where $\partial_n$ denotes the outward facing normal derivative. The case $\eps\ll1$ in which $|\partial \Omega_\eps| \ll |\partial \Omega|$ is known as the narrow escape or narrow capture problem \cite{HolcmanReview2014,Holcman2007PNAS} and signifies the scenario where the extent of the absorbing set is significantly less than the total search domain. In \eqref{MainFreeProb_a}, we have set the diffusion coefficient to unity without loss of generality. The goal is to obtain the free probability $P(t; \bxi)$ \jt{(often referred to as the complementary cumulative distribution of the capture time $t$)} and the capture time density $C(t; \bxi)$
\begin{equation} \label{freedens}
P(t;\bxi) = \int_{\Omega\setminus\Omega_{\eps}} \! p(\bx,t;\bxi) \, d \bx \,, \qquad C(t; \bxi) = -\frac{dP}{dt} \,,
\end{equation}
where $p(\bx,t;\bxi)$ solves \eqref{MainFreeProb}. 


Partially motivated by the large number of individual walkers in applications, the mean first passage time (MFPT) is the commonly studied first moment of $C(t; \bxi)$. The MFPT $w(\bx)$ of a particle starting from $\bx\in\Omega$ satisfies the simpler elliptic boundary value problem of mixed Dirichlet-Neumann type (cf. \cite{redner2001guide})
\bsub \label{mfpteq_intro}
\BE \label{mfpteq_intro_a}
\Delta w = -1 \, , \quad \bx\in\Omega\setminus\Omega_{\eps}; \qquad \partial_n w = 0\,, \quad x \in \partial \Omega \,, \qquad w = 0 \,, \quad x \in \partial \Omega_\eps \,.
\EE
Equation \eqref{mfpteq_intro} has been studied (cf. \cite{HolcmanReview2014,redner2001guide,Holcman2007PNAS} and references therein) for a variety of absorbing sets such as internal traps \cite{O,KTW,LTK2015}, boundary escape windows \cite{singer2006narrow,Pillay2010} and dumbbell domains with narrow necks \cite{HolcmanReview2014}. In three dimensions, equation \eqref{mfpteq_intro} has been studied in the spherical case for internal and boundary windows \cite{ChevWard2010,Ward2010,IN2013,Coombs2009,ChevZawada2013,Holcman2006a} and non-spherical geometries with absorbing boundary windows \cite{HolcmanLeakage2008,Cheviakov2015}. \al{A common measure of the capture rate is given by the global MFPT which, for a uniform distribution $\rho(\bx) = |\Omega\setminus\Omega_{\eps}|^{-1}$ of initial walker locations, is given by}
\begin{equation} \label{mfpteq_intro_b}
\tau = \int_{\Omega\setminus\Omega_{\eps}} \rho(\bx) w(\bx)d\bx = \frac{1}{|\Omega\setminus\Omega_{\eps}|} \int_{\Omega\setminus\Omega_{\eps}} w(\bx)d\bx.
\end{equation}
\esub  
The quantity $\tau$ gives the broadest possible measure of the capture rate by the absorbing set $\partial\Omega_{\eps}$.

\begin{figure}[htbp]
  \centering
    \includegraphics[width=0.45\textwidth]{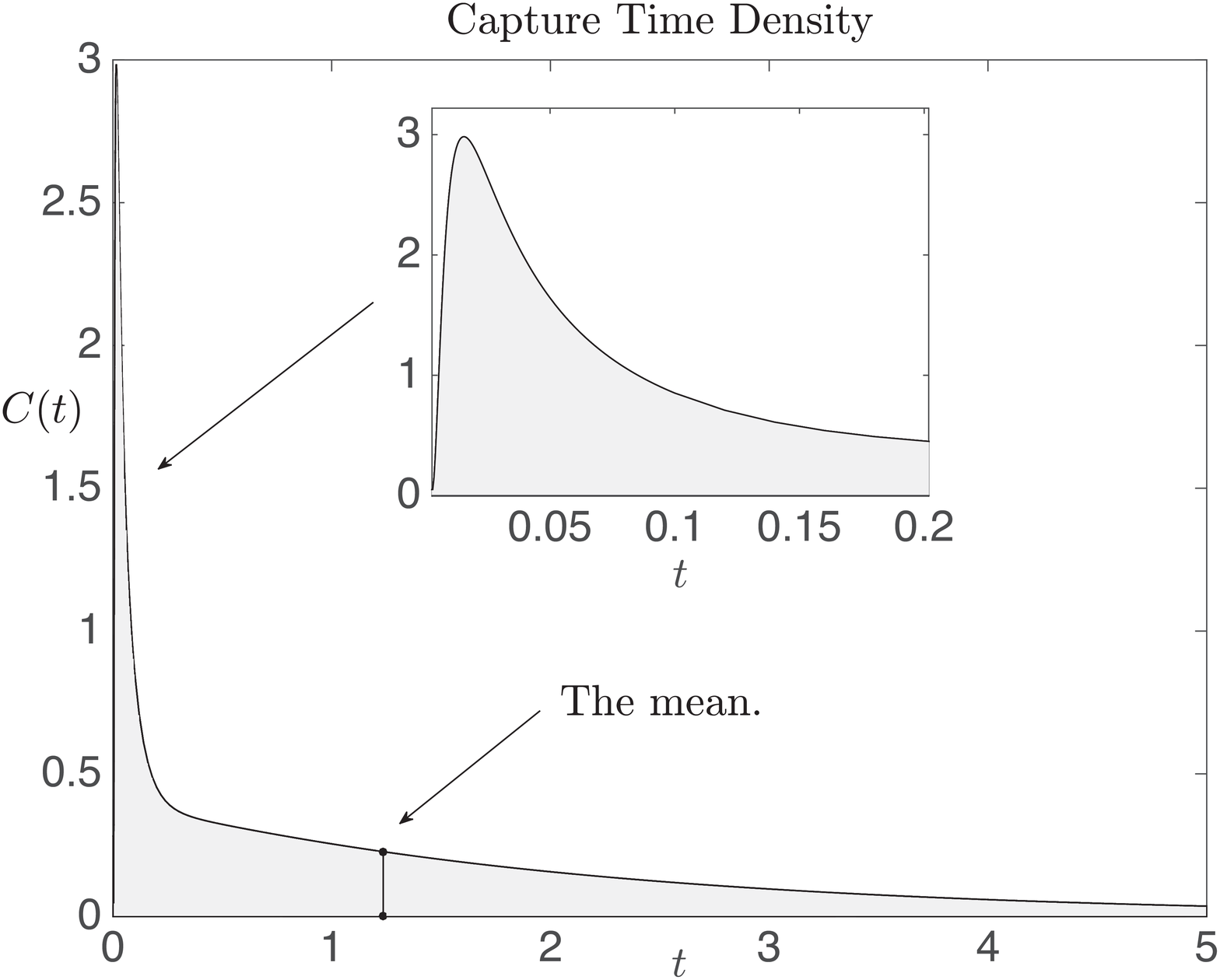}
      \parbox{0.75\textwidth}{\caption{The full distribution of capture times $C(t)$ for a random walker in the unit disk, starting at $(0.3,0)$, and absorption at a circular trap of radius $\eps=0.01$ centered at the origin. \label{Fig:CaptureDensity}}}
\end{figure}

The narrow escape problem is a rare event process and as such passage times may be broadly distributed around the MFPT. For this reason, the MFPT does not necessarily reflect typical capture times which limits the usefulness of problem \eqref{mfpteq_intro} in describing the underlying stochastic process. The goal of this paper is to present a methodology for calculating the full passage time density in two dimensional domains with multiple small non-overlapping internal traps. To illustrate our method, we plot in Fig.~\ref{Fig:CaptureDensity} the distribution of absorption times to a single small internal target obtained by our method applied to \eqref{MainFreeProb} and compare with the MFPT obtained from \eqref{mfpteq_intro}.

There are two important details which are apparent from Fig.~\ref{Fig:CaptureDensity}. First, the distribution has a pronounced peak at short times $t$. This \jt{peak} is a factor of the geometry and the initial position of the walker, which is not captured by the MFPT, and reflects the Brownian paths which are quickly absorbed by the trap. Second, the capture time is broadly distributed about the mean and exhibits a long flat tail implying that many Brownian paths have large excursions before eventual capture.  

To elucidate the second of these points, for the case of a single absorbing trap of radius $\eps$ centered at $\bx_0$, we calculate in \S \ref{sec:variance} the variance $\mathcal{V}(\bx)$ and standard deviation $\sigma(\bx)=\sqrt{\mathcal{V}(\bx)}$ of the passage time as $\eps\to0$. A key component in this calculation, is knowledge of the Neumann's Green's function $G_m(\bx;\bxi)$ and its regular part $R_m(\bx;\bx_0)$ satisfying
\bsub\label{modGreens}
\begin{gather}
\Delta G_{m}=\frac{1}{|\Omega|}-\delta(\bx - {\bs \xi}),\quad \bx\in\Omega;\qquad
\partial_n G_{m} =0,\quad \bx\in\partial\Omega
;\label{modGreens_b}\\[5pt]
\int_{\Omega}G_m(\bx;\bxi)\,\mathrm{d}\bx=0;\qquad G_m(\bx;\bxi)=\frac{-1}{2\pi}\log
|\bx - {\bs \xi}|+R_{m}(\bx;{\bs \xi}).
\end{gather}
\esub
In \cite{Ward2010} it was shown for the single trap case $\Omega_{\eps}= \bx_0 + \eps\Omega_0$, that the MFPT satisfies $w(\bx) = w_0(\bx;\nu) + \bigoh(\eps)$ where
\bsub\label{intro:MFPT}
\begin{equation}\label{intro:MFPT_a}
w_0(\bx) =  -|\Omega| G_m(\bx;\bx_0) + \chi_0.
\end{equation} 
Here $\chi_0 = |\Omega|^{-1}\int_{\Omega}w_0(\bx)d\bx$ is the global MFPT averaged over a uniform distribution of \jt{starting} locations given by
\begin{equation}\label{intro:MFPT_b}
\chi_0 = \frac{|\Omega|}{2\pi} \left[ \frac{1}{\nu} + 2\pi R_m(\bx_0;\bx_0)\right], \qquad \nu = \frac{-1}{\log \eps d_0},
\end{equation}
\esub
and $d_0$ is the capacitance of the trap, determined from the exterior problem \eqref{innerc}. In \S \ref{sec:variance} we calculate as $\eps\to0$ the variance $\mathcal{V}(\bx)=\mathcal{V}_0(\bx) +\bigoh(\eps)$  where 
\bsub\label{intro:var}
\begin{equation}\label{intro:var_a}
\mathcal{V}_0(\bx) = \chi_0^2 -|\Omega|^2\left( [G_m(\bx;\bx_0)]^2 - \frac{2}{|\Omega|} \int_{\Omega}([G_m(\by;\bx_0)]^2-G_m(\by;\bx)G_m(\by;\bx_0)  ) \, d\by \right).
\end{equation}
An inspection of \eqref{intro:var_a} reveals that the standard deviation $\sigma(\bx) = \sqrt{\mathcal{V}_0(\bx)}$ is asymptotically equal to the mean, up to two terms in $\nu$ as $\eps\to0$. Moreover, the standard deviation depends on the initial location $\bx\in\Omega$ only in the third term of its expansion. This weak dependence indicates that even walkers which start close to the target are not necessarily captured quickly and typically undergo long excursions before absorption. Averaging over a uniform distribution of initial locations, the global variance of the passage times is calculated to be
\begin{equation}\label{intro:var_b}
\frac{1}{|\Omega|} \int_{\Omega} \mathcal{V}_0(\bx) d \bx = \chi_0^2 + |\Omega|\int_{\Omega} [G_m(\bx;\bx_0)]^2 d\bx.
\end{equation}
\esub
Therefore, in the limit as $\eps\to0$, the global standard deviation $(|\Omega|^{-1} \int_{\Omega} \mathcal{V}_0(\bx) d \bx)^{\frac12}$ is strictly greater than the global MFPT indicating that the mean may not be a reliable estimator of the capture time.

For this reason, we are motivated in \S \ref{sec:FullDist} to investigate a methodology for obtaining the full distribution of passage times by constructing solutions to \eqref{MainFreeProb} in the limit as $\eps\to0$. The solution to \eqref{MainFreeProb} may be expressed in terms of a separable solution:
\begin{equation}\label{eq:expansion_prob}
p(\bx,t) = \sum_{n=0}^{\infty} c_n e^{-\lambda_n t}\phi_n(\bx), \qquad c_n = \int_{\Omega} \phi_n(\bx) \delta(\bx-\bxi)d\bx = \phi_n(\bxi),
\end{equation}
where the Laplacian eigenfunctions $\phi_n(\bx)$ and eigenvalues $\lambda_n$ satisfy
\begin{equation}\label{eigproblem}
\Delta \phi + \lambda \phi = 0, \quad \bx\in\Omega; \qquad \phi = 0, \quad \bx\in\partial\Omega_{\eps}, \qquad \partial_n \phi = 0, \quad \bx\in\partial\Omega,  \qquad \int_{\Omega\setminus\Omega_{\eps}} \phi^2\, d\bx =1.
\end{equation}
The ordering principle of Laplacian eigenvalues $0<\lambda_0 <\lambda_1 \leq \lambda_2  \leq \lambda_3 \cdots$ indicates that for large $t$, the density behaves asymptotically as\begin{equation}\label{LargeTimeApprox}
p(\bx,t) \sim c_0 e^{-\lambda_0(\eps) t}\phi_0(\bx),\qquad \mbox{as} \qquad t\to\infty.
\end{equation}
Therefore the free probability is exponentially decaying with rate largely dependent on the \jt{principal} eigenvalue $\lambda_0(\eps)$. The value of $\lambda_0(\eps)$ has been accurately estimated in the limit as $\eps\to0$ for many narrow escape problems \cite{F,KTW,O,WHK,WK,Pillay2010,Ward2010,LTK2015}. However, the expansion solution \eqref{eq:expansion_prob} is not informative for small values of $t$ where all terms in the expansion \eqref{eq:expansion_prob} make a contribution. Moreover, as seen in Fig.~\ref{Fig:CaptureDensity}, the capture time distribution has important features such as a prominent peak (or multiple peaks) which occurs at short times and are not described by the monotonic large time approximation \eqref{LargeTimeApprox}. Therefore, the main missing piece in the analysis of the capture time density is determination of the short time behavior.

The problem of obtaining the full distribution of capture times has been considered for certain special regions such as circular wedges \cite{Silva09Wedge,OshaninWedges2012} and radially symmetric domains \cite{benichou2014first, godec2016first}. In these cases, the absorbing set is placed strategically to preserve radial symmetry, allowing for exact solutions in the forms of infinite series. For a spherical domain with a single interior trap, \cite{IN2013} employed a pseudopotential method to develop a small time correction to the large time approximation \eqref{LargeTimeApprox}. In the present work, we develop a methodology for obtaining the distribution in a general two dimensional region with multiple non-overlapping traps. To accommodate the time dependent nature of \eqref{MainFreeProb}, we apply the Laplace transform (cf. \cite{redner2001guide,benichou2014first}) which is commonly used in determining short time behavior of parabolic problems. The resulting equation for the Laplace transform $u(\bx;s,\bxi)$ is elliptic and of modified Helmholtz type with small perturbing holes;
\bsub\label{intro_LT}
\begin{gather}
\label{intro_LTa} \Delta u - su = - \delta(\bx - \bxi)\,, \quad \bx\in\Omega\,; \\[5pt]
\label{intro_LTb} \partial_n u = 0\,, \quad \bx\in\partial\Omega\,; \qquad u = 0\,, \quad \bx \in \partial\Omega_{\eps}\,.
 \end{gather}
 \esub
 To solve equation \eqref{intro_LT}, we apply the method of matched asymptotic expansions in the limit of small trap size $\eps\to0$, and determine a solution of \eqref{intro_LT} in terms of the modified Helmholtz Green's function
 \bsub\label{helmholtzGreens}
\begin{gather}
\label{helmholtzGreens_a} \Delta G_h - \lambda^2 G_h = -\delta(\bx - \bxi), \quad \bx\in\Omega; \qquad \partial_n G_h = 0, \quad \bx\in \partial\Omega \,;\\[5pt]
\label{helmholtzGreens_b} G_h(\bx;\bxi,\lambda^2) \sim \frac{-1}{2\pi} \log|\bx -\bxi| + R_h(\bxi;\bxi,\lambda^2) \quad \mbox{as} \quad \bx \to \bxi \,, \qquad \int_{\Omega}G_h(\bx;\bxi,\lambda^2)\, d\bx = \frac{1}{\lambda^2}.
\end{gather}
\esub
  In certain limits, such as $t\to0$, we are able to obtain an asymptotic solution for \eqref{helmholtzGreens}, while in others we utilize a finite element solver \cite{supp}. To invert the Laplace transform back to the time domain, we use a numerical inversion technique \jt{\cite{Abate2006,mcclure}}. We emphasize that the hybrid technique we present in \S \ref{sec:FullDist} - \S \ref{larget} requires only solutions of time-independent problems, \jt{is} valid for arbitrary domain and trap geometries, and \jt{estimates} $C(t)$ beyond all orders in $\nu \equiv \mathcal{O}( -1/\log\eps)$.

\setcounter{equation}{0}
\section{The variance of the first passage time}\label{sec:variance}

In this section, we calculate the variance $\mathcal{V}(\bx) = \mathbb{T}(\bx)  - w(\bx)^2$ and standard deviation $\sigma(\bx) =\sqrt{\mathcal{V}(\bx)}$ where $\mathbb{T}(\bx)$ is the second moment of the distribution and satisfies the elliptic problem (cf. \cite{delgado2015conditional})
\begin{equation}\label{VarMain}
\Delta \mathbb{T} = -2w(\bx), \quad \bx\in\Omega\setminus\Omega_{\eps}; \qquad \partial_n \mathbb{T} = 0, \quad \bx\in\partial\Omega, \qquad \mathbb{T} = 0, \quad \bx\in \partial\Omega_{\eps}.
\end{equation}
In the limit as $\eps\to0$, we develop a solution to \eqref{VarMain} with a single trap expressed as $\Omega_{\eps} = \bx_0 + \eps \Omega_0$. This analysis can be extended to multiple traps (cf. \cite{Kurella2015}). The solution is expanded in the form
\[
\mathbb{T}(\bx)  = \mathbb{T}_0(\bx;\nu)  + \eps \mathbb{T}_1(\bx;\nu) + \cdots, \qquad \nu = \frac{-1}{\log\eps d_0}.
\]
We are interested in the leading order problem for $\mathbb{T}_0(\bx)$ which satisfies 
\bsub\label{eqnT0}
\begin{equation}\label{eqnT0_a}
\Delta \mathbb{T}_0 = -2w_0(\bx), \quad \bx\in\Omega\setminus\{ \bx_0\}; \qquad \partial_n \mathbb{T} = 0, \quad \bx\in\partial\Omega.
\end{equation}
The local condition on $\mathbb{T}_0(\bx)$ as $\bx\to\bx_0$ is found to be
\begin{equation}\label{eqnT0_b}
\mathbb{T}_0(\bx) \sim A \log|\bx-\bx_0| + \frac{A}{\nu}, \qquad \bx\to\bx_0; \qquad \nu = \frac{-1}{\log\eps d_0},
\end{equation}
\esub
where $A$ is the strength of the singularity and will be determined from a solvability condition. The logarithmic capacitance $d_0$ is a trap parameter determined from \al{ a local problem in the vicinity of $\bx_0$ where $\mathbb{T}(\bx_0 + \eps \by)= v_c(\by)$ and $\by$ is the stretched variable $\by = (\bx-\bx_0)/\eps$. The problem for $v_c(\by)$ is}
\bsub\label{innerc}
\begin{gather}
\Delta v_{c}=0,\qquad \by\in\mathbb{R}^{2}\setminus\Omega_{0}; \qquad v_{c}=0\,\quad\mbox{on}\quad\Gamma_{0}^{a},\qquad\frac{\partial v_{c}%
}{\partial n}=0\quad\mbox{on}\quad\Gamma_{0}^{r};\\
v_{c}(\by)=\log|\by|-\log d_{0}+\frac{\mathbf{p}\cdot \by}{|\by|^{2}}%
+\bigoh\Big(\frac{1}{|\by|^{2}}\Big),\qquad|\by|\rightarrow\infty, \label{innerc_d}
\end{gather}
\esub
The value of $d_0$ depends on the shape of the rescaled trap $\Omega_0$ and its distribution of absorbing and reflecting portions, $\Gamma_0^a$ and $\Gamma_0^r$, respectively. For all absorbing traps, the value of $d_0$ can be calculated for a variety of regular (circles, ellipses, triangles) shapes \cite{Kurella2015}. When $\Omega_0$ is the unit disk, $d_0 = 1$.  When the trap is not uniformly absorbing, $d_0$ has been calculated for a variety of configurations and homogenized limits \cite{LTK2015}. The vector $\mathbf{p}$ is the dipole and depends on the orientation of the trap - its influence on the narrow escape problem was analyzed in \cite{LTK2015}. If a Robin condition $\partial_n v_c + \kappa v_c = 0$ is applied to a circular trap $\partial\Omega$ of unit radius, then the capacitance is $d_0 = e^{1/\kappa}$ \jt{\cite{LTK2015}.}

The local condition \eqref{eqnT0_b} gives rise to a Dirac source term on the right hand side of \eqref{eqnT0_a}. Substituting \eqref{intro:MFPT_a} into \eqref{eqnT0_a} gives
\begin{equation}\label{eqn2T0_a}
\Delta \mathbb{T}_0 =  2|\Omega| G_m(\bx;\bx_0)  - 2\chi_0 + 2\pi A\delta(\bx-\bx_0), \quad \bx\in\Omega; \qquad \partial_n \mathbb{T} = 0, \quad \bx\in\partial\Omega.
\end{equation}
Integrating \eqref{eqn2T0_a} over $\Omega$ and applying the divergence theorem yields the solvability condition 
\begin{equation}\label{eqn:Source}
A = \frac{|\Omega|\chi_0}{\pi}.
\end{equation}
To obtain a solution to \eqref{eqnT0_a}, we first decompose $\mathbb{T}_0 = \mathbb{T}_{p} + \mathbb{T}_{h}$ where
\bsub\label{eqns:T}
\begin{gather}
\label{eqn:Th} \Delta \mathbb{T}_{h} = -2|\Omega|\chi_0\left[ \frac{1}{|\Omega|} -  \delta(\bx-\bx_0) \right], \quad \bx\in\Omega;  \qquad  \partial_n \mathbb{T}_h = 0, \quad \bx\in\partial\Omega;\\[5pt]
\label{eqn:Tp} \Delta \mathbb{T}_{p} =   2|\Omega|G_m(\bx;\bx_0), \quad \bx\in\Omega;  \qquad  \partial_n \mathbb{T}_p = 0, \quad \bx\in\partial\Omega, \qquad \int_{\Omega}\mathbb{T}_p(\bx)\, d\bx =0.
\end{gather}
\esub
The solutions of these problems are given by
\begin{gather}
 \mathbb{T}_{h}(\bx) = -2|\Omega|\chi_0 G_m(\bx;\bx_0) + \chi_1, \qquad \mathbb{T}_{p}(\bx) = -2|\Omega| \int_{\Omega}G_m(\by;\bx) G_m(\by;\bx_0)\, d\by.
\end{gather}
The final unknown $\chi_1 = |\Omega|^{-1}\int_{\Omega}\mathbb{T}_0\, d \bx$ is fixed by matching $\mathbb{T}_0 = \mathbb{T}_{p} + \mathbb{T}_{h}$ to the local condition \eqref{eqnT0_b} as $\bx\to\bx_0$ which yields that
\[
\frac{|\Omega|\chi_0}{\nu\pi} = \mathbb{T}_p(\bx_0) - 2|\Omega|\chi_0 R_m(\bx_0;\bx_0) + \chi_1.
\]
Rearranging for $\chi_1$ and recalling the definition of $\chi_0$ in \eqref{intro:MFPT_b} gives
\begin{equation}
\chi_1 = 2\chi_0^2 - \mathbb{T}_p(\bx_0).
\end{equation}
The leading order variance $\mathcal{V}_0(\bx) = \mathbb{T}_0(\bx) - w_0(\bx)^2$ can now be calculated as 
\begin{align}
\nonumber \mathcal{V}_0(\bx) & = \left[ \mathbb{T}_p(\bx) - 2|\Omega| \chi_0G_m(\bx;\bx_0) + 2\chi_0^2 - \mathbb{T}_p(\bx_0)\right] - \left[  -|\Omega| G_m(\bx;\bx_0) + \chi_0\right]^2\\[5pt]
\nonumber {} & = \chi_0^2 - \left( |\Omega|^2[G_m(\bx;\bx_0)]^2 + \mathbb{T}_p(\bx_0) -  \mathbb{T}_p(\bx)\right)\\[5pt]
\label{V0Full} {} & = \chi_0^2 - |\Omega|^2\left( [G_m(\bx;\bx_0)]^2 - \frac{2}{|\Omega|} \int_{\Omega}([G_m(\by;\bx_0)]^2-G_m(\by;\bx)G_m(\by;\bx_0)  ) \, d\by \right).
\end{align}
Recalling that $\chi_0$ is the global MFPT \eqref{intro:MFPT_b} and $\sigma \sim \sigma_0 = \sqrt{\mathcal{V}_0}$, we obtain the behaviors
\bsub\label{LOVarSig}
\begin{align}
\label{LOVar} \mathcal{V}_0(\bx) &= \frac{|\Omega|^2}{4\pi^2 \nu^2}\Big[ 1 +4\pi \nu R_m(\bx_0;\bx_0)\Big] + \bigoh(1), \quad \mbox{as} \quad \eps\to0,\\[5pt]
\label{LOSig} \mathcal{\sigma}_0(\bx) &= \frac{|\Omega|}{2\pi }\Big[ \ \frac{1}{\nu} + 2\pi R_m(\bx_0;\bx_0) \Big] + \bigoh(\nu) , \quad \mbox{as} \quad \eps\to0.
\end{align}
\esub
There are two important implications of this result. First, the MFPT and standard deviation of mean are asymptotically equal up to two terms as $\eps\to0$. Therefore, the MFPT may not necessarily be a reliable measure since the distribution of passages times around the mean is relatively broad. This is the principal motivation for developing (cf.~\S \ref{sec:FullDist}) a methodology for obtaining the full capture time density. 

The second implication from \eqref{LOSig} is that as $\eps\to0$, the standard deviation $\sigma(\bx)$ largely depends on the trap location $\bx_0$ and is weakly dependent on the starting location $\bx$. From \eqref{intro:MFPT_a}, we see that if $\bx$ is close to $\bx_0$, then since $G_m(\bx;\bx_0)>0$, the mean capture time is less than the global MFPT $\chi_0$. However, in \eqref{LOSig}, the standard deviation is unchanged, up to two orders, as $\bx$ approaches $\bx_0$. The interpretation is that even if the Brownian walker starts close to the trap, there are still many random paths which are not captured quickly and undergo large excursions in $\Omega$ before eventual absorption. \jt{While they may be rare, the duration of these excursions may be asymptotically long and are the source of the $\bigoh(\nu^{-1})$ mean capture time. This leads to a large disparity between the mean and the mode of the capture time, the latter of which may often (depending on the specific application) be a more informative measure of first passage processes.}

We also remark that the asymptotic expressions for the global MFPT and the standard deviation $\sigma_0$, given in \eqref{intro:MFPT_b} and \eqref{LOSig}, respectively, are both minimized when the trap location $\bx_0$ minimizes $R_m(\bx_0;\bx_0)$. For a uniform distribution of initial locations, the global variance of the first passage time can be calculated as 
\begin{equation}\label{GlobalVPT}
\frac{1}{|\Omega|} \int_{\Omega} \mathcal{V}_0(\bx) d \bx = \al{\sigma_0^2(\bx)}= \chi_0^2 + |\Omega|\int_{\Omega} [G_m(\bx;\bx_0)]^2 d\bx,
\end{equation}
indicating that the global standard deviation is strictly larger than the global MFPT. The global coefficient of variation $c_{V}$ is an important measure of variability in the distribution defined as the ratio of the global standard deviation \eqref{GlobalVPT} to the global MFPT \eqref{intro:MFPT_b}. We calculate that
\[
c_{V} = \al{\frac{\sigma_0(\bx)}{\chi_0}} = \sqrt{1 + \frac{|\Omega|}{ \chi_0^2}\int_{\Omega} [G_m(\bx;\bx_0)]^2d\bx } >1
\]
This further reinforces the notion that the MFPT can be an unreliable estimator of the capture time.

As a confirmation of \eqref{V0Full} and its two-term approximation \eqref{LOVarSig}, we compare against the exactly solvable situation of a circular trap of radius $\eps$ centered at the origin of the unit disk $\Omega = \{ \bx \in \mathbb{R}^2 \  |\ \eps< |\bx| <1 \}$. The exact solutions of \eqref{mfpteq_intro} and \eqref{VarMain} in terms of the radial variable $r=|\bx|$ are given by
\bsub\label{ExactDisk}
\begin{align}
\label{ExactDisk_a} w(r) &= \frac{-1}{4} \left[r^2-\eps^2 + 2 \log\frac{r}{\eps} \right],\\[5pt]
\label{ExactDisk_b} \mathbb{T}(r) &= \frac{1}{32} \left[r^4 + 8(r^2-\eps^2) - 4r^2\eps^2 + 3 \eps^4  -12\log\frac{r}{\eps} - 8(r^2-\eps^2)\log\frac{r}{\eps} - 16\log\eps\log\frac{r}{\eps}\right] \,.
\end{align}
\esub
As $\eps\to0$, we obtain the leading order solutions
\bsub\label{ExactDiskLim}
\begin{align}
\label{ExactDiskLim_a} w_0(r) &= \frac{1}{2} \left[ \frac{1}{\nu}  +  \log r - \frac{r^2}{2}\right] + \bigoh(\eps^2),\\[5pt]
\label{ExactDiskLim_b} \mathbb{T}_0(r) &= \frac{1}{32} \left[\frac{16}{\nu^2} -\frac{4}{\nu} \left( 3+ 2r^2 - 4\log r\right) + r^4 + 8r^2 -4(3+2r^2)\log r \right] + \bigoh(\eps^2\log\eps),\\[5pt]
\label{ExactDiskLim_c} \mathcal{V}_0(r) &= \frac{1}{4} \left[ \frac{1}{ \nu^2}  - \frac{3}{2\nu} + \left( r^2 - \frac{r^4}{8} - \frac{3}{2} \log r - \log^2 r\right) \right] + \bigoh(\eps^2\log\eps).
\end{align}
\esub
To compare with the asymptotic solution, we use the explicit expression for the Neumann's Green's function for the unit disk (cf.~{\cite{KTW})
\bsub \label{GRmdisk_Exact}
\BE \label{Gmdisk_Exact}
	G_m(\bx;\bxi) = \frac{1}{2\pi} \left( -\log |\bx - \bxi| - \log\left| \bx|\bxi| - \frac{\bxi}{|\bxi|} \right|+ \frac{1}{2}\left(|\bx|^2 + |\bxi|^2 \right) - \frac{3}{4} \right) \,,
\EE
\BE \label{Rmdisk_Exact}
  R_m(\bxi;\bxi) =  \frac{1}{2\pi} \left( - \log\left| \bxi|\bxi| - \frac{\bxi}{|\bxi|} \right|+ |\bxi|^2 - \frac{3}{4} \right) \,.
\EE
\esub
Setting $\bx = (r,0)$, we have that $2\pi R_m({\bf 0},{\bf 0})= -3/4$ and $2\pi G_m(\bx,{\bf 0}) = -\log r + r^2/2 -3/4$ so that \eqref{intro:MFPT} and \eqref{ExactDiskLim_a} are in agreement. By direct calculation, $\int_{\Omega} [G_m(\bx;\bx_0)]^2d\bx = 7/192\pi$, so that \eqref{GlobalVPT} yields
\begin{equation}\label{GlobalVPT_Disk} 
\frac{1}{|\Omega|} \int_{\Omega} \mathcal{V}_0(\bx)\, d \bx = \frac{1}{4}\left[ \frac{1}{\nu^2} - \frac{3}{2\nu} + \frac{17}{24} \right].
\end{equation}
From the exactly solvable case, we calculate using \eqref{ExactDiskLim_c} that
\begin{equation}\label{GlobalVPT_Disk_a}
\frac{1}{|\Omega|} \int_{\Omega} \mathcal{V}_0(\bx)\, d \bx = \frac{1}{\pi} \int_{\theta =0}^{2\pi} d\theta \int_{r=\eps}^{1} \mathcal{V}_0(r)\, r dr = \frac{1}{4}\left[ \frac{1}{\nu^2} - \frac{3}{2\nu} + \frac{17}{24} \right] + \bigoh(\eps^2\log\eps).
\end{equation}
in complete agreement with \eqref{GlobalVPT_Disk}.

\setcounter{equation}{0}
\section{Determination of the full capture time distribution}\label{sec:FullDist}

In the analysis below, we develop a methodology for obtaining the full distribution of capture times in general two dimensional domain with small traps. There have been many recent treatments focussed solely on determination of the MFPT in the well-known narrow escape problem in two and three dimensions (see, e.g., \cite{Holcman2006a, Holcman2006b, Coombs2009, Pillay2010, Ward2010}). The mean first passage time (MFPT) $w(\bx)$ and the second moment of the FPT $\mathbb{T}(\bx)$ starting from $\bx$ can be obtained from the capture time density
\BE \label{mfpt}
w(\bx) = \int_{0}^\infty \! t C(t; \bx) \, dt \,, \qquad \mathbb{T}(\bx) = \int_{0}^\infty \! t^2 C(t; \bx) \, dt \,.
\EE

%
%
%
%
%

To calculate $P$ and $C$, we start by introducing the Laplace transform $u(\bx,s;\bx_0)$ of \eqref{MainFreeProb} defined by
\begin{equation}\label{LaplaceTransform}
u(\bx,s;\bx_0)=\mathcal{L}[p](s) = \int_0^{\infty}\! e^{-st} p(\bx,t;\bx_0) \, dt \,.
\end{equation}

From \eqref{MainFreeProb}, the equation satisfied by $u(\bx,s;\bx_0)$ is then
\bsub\label{LT1}
\begin{gather}
\label{LT1a} \Delta u - su = - \delta(\bx - \bx_0)\,, \quad \bx\in\Omega\,; \\[5pt]
\label{LT1b} \partial_n u = 0\,, \quad \bx\in\partial\Omega\,; \qquad u = 0\,, \quad \bx \in \partial\Omega_{\eps}\,.
 \end{gather}
 \esub
 The free probability density $P(t)$ then satisfies
 \bsub \label{LTofP}
 \BE 
 	\mathcal{L}\lbrack P \rbrack(s) =  U(\bx_0,s)  \,, \qquad \mathcal{L}\lbrack C \rbrack(s) =  -s U(\bx_0,s) + P(0, \bx_0) \,,
 \EE
 where $U(\bx_0, s)$ is given by
 \BE
 U(\bx_0, s) = \int_{\Omega\setminus\Omega_{\eps}} \! u(\bx,s;\bx_0)\,d \bx \,.
 \EE
 \esub
 Hence, the Laplace transform of the capture time density satisfies an elliptic PDE \eqref{LT1} of modified Helmholtz type with mostly reflecting boundaries and small Dirichlet portions representing absorbing traps. In the next section\jt{,} we develop solutions to \eqref{LT1} in the limit as $\eps\to0$ \jt{and utilize a numerical inverse Laplace transform} to obtain the full capture time density.
 

\subsection{Solution in two dimensions with interior traps}
In this section we outline a hybrid asymptotic-numerical method, based on matched asymptotic expansions, for computing the solution of \eqref{LT1} in two dimensions with $N$ non-overlapping internal traps of size $\bigoh(\eps)$. These traps occupy the regions $\Omega_{\eps_j} = \bx_j + \eps\, \Omega_j$ with trap centers $\bx_j \in\Omega$ for $j = 1,\ldots,N$.

As is common in the asymptotic analysis of two dimensional problems with small domains removed \cite{Pillay2010,WK,KTW,LTK2015}, each trap may be replaced as $\eps\to0$ by a logarithmic singularity of prescribed strength $A_j(s)$ with an associated regular part
 \begin{equation}\label{SingStrength}
 u(\bx) \sim A_j(s) \nu_j \log|\bx - \bx_j| + A_j(s) + \cdots \quad \bx\to \bx_j, \qquad j = 1,\ldots,N; \qquad \nu_j \equiv \frac{-1}{\log \eps d_j} \,.
 \end{equation}
The parameter $d_j$ is the logarithmic capacitance of the $j$-th trap and is determined from \eqref{innerc}. In terms of the modified Helmholtz Green's function $G_h(\bx;\bxi,\lambda^2)$ satisfying  \eqref{helmholtzGreens}, the solution of \eqref{LT1} with specified singularity behavior \eqref{SingStrength} is
\begin{equation}\label{LT2}
u(\bx;\bx_0,s)  = G_h(\bx;\bx_0,s) - 2\pi \sum_{j=1}^{N}  A_j \nu_jG_h(\bx; \bx_j, s) \,.
\end{equation}
A system of equations for the trap strengths $A_j$, which indicate the flux over each trap, is obtained by matching \eqref{LT2} to the singularity behavior \eqref{SingStrength} as $\bx\to\bx_k$ for each $k=1,\ldots,N$. This generates the system of linear equations
\BE \label{AkRk}
 A_k = G_{k,0} - 2\pi \left[ A_k \nu_kR_{k,k} + \sum_{j\neq k}^N A_{j}\nu_jG_{k,j} \right], \quad k = 1,\ldots N; \qquad \left\{ \begin{array}{rcl} R_{k,k} = R_h(\bx_k;\bx_k,s) \\[5pt] G_{j,k} = G_h(\bx_j;\bx_k,s)\end{array}\right. \,,
\EE
 which can be concisely represented in matrix form as
 \bsub\label{LinA}
 \begin{equation} \label{AkRkmat}
 [I + 2\pi \, \mathcal{G} \, V  ] \bA = \bg_0, \qquad 
 \end{equation}
 where we have defined
\begin{equation} \label{Gmat}
 \begin{array}{l} \bA= [A_1, \ldots, A_N]^{T}\\[5pt] \bg_0 = [G_{1,0}, \ldots G_{N,0}]^{T} \end{array}, \qquad \mathcal{G}_{i,j} = \left\{ \begin{array}{rcl} R_{i,i} &\mbox{if}& i = j \\[5pt] G_{i,j} &\mbox{if} & i\neq j \end{array}\right. \,, \qquad V_{i,j} = \left\{ \begin{array}{rcl} \nu_i &\mbox{if}& i = j \\[5pt] 0&\mbox{if} & i\neq j \end{array}\right. \,.
 \end{equation}
 \esub
 With \eqref{helmholtzGreens_b} and \eqref{LT2} along with $U(\bx,s)$ defined in \eqref{LTofP}, we have that in the limit as $\eps\to0$, the free probability $P(t)$ satisfies
 \bsub \label{LPC}
 \begin{equation} \label{LP}
\mathcal{L}[P](s) = U(\bx_0,s) = \frac{1}{s}\left[ 1 - 2\pi\sum_{j=1}^{N} \nu_jA_j(s) \right] \,.
 \end{equation}
The capture time density $C(t;\bx_0)$ defined in \eqref{freedens} then satisfies
\begin{equation} \label{LC}
\mathcal{L}[C](s) = -s\mathcal{L}[P](s) + P(0;\bx_0) = 2\pi \sum_{j=1}^{N} \nu_jA_j(s) \,,
\end{equation}
\esub
 where we have used that $P(0;\bx_0) = 1$. In \eqref{LC}, the dependence of the capture rate $C(t)$ on the initial location of the walker $\bx_0$ and the trap centers $\{\bx_1,\ldots, \bx_N\}$ is encoded in the strengths $A_k$ satisfying the linear system \eqref{LinA}. Using this matched asymptotic method, the increase in complexity introduced by each additional trap is simply  accounted for by an extra dimension in the linear system \eqref{AkRkmat}. The associated error, following the analysis of \cite{Ward2000}, can be shown to be smaller than $\max_j\nu_j^m$ for any integer $m$.
 
The key to obtaining $P(t;\bx_0)$ and $C(t;\bx_0)$ lies in being able to accurately compute the modified Helmholtz Green's function $G_h(\bx;\bxi;\lambda)$ in \eqref{helmholtzGreens}. When $\Omega$ is the unit square, $G_h$ may be expressed analytically in terms of an infinite series using the method of images \cite{kolokolnikov2003reduced}. When $\Omega$ is the unit circle, $G_h$ may be found in terms of an infinite Fourier series. Efficient numerical solution of the modified Helmholtz equation is a problem with a rich history in acoustics and many well developed  solution methodologies exist based on finite difference, finite element \cite{Harari2006}, integral function \cite{Kropinski2011,tai1974helmholtz}, and spectral  \cite{Smitheman2010,Cheng2006} techniques. 
 
 An important step in numerical or analytical consideration of \eqref{helmholtzGreens} is a separation of the regular and singular parts of $G_h$ 
 \BE \label{GRV}
 	G_h(\bx;\bxi,\lambda^2) =  V(\bx;\bxi,\lambda^2) +\tilde{R}(\bx;\bxi,\lambda^2) \,,
 \EE
 \noindent where $V(\bx;\bxi,\lambda^2)$ is the free-space Green's function. In spatial dimension $d=2$, 
\bsub
 \begin{gather}
  \label{Veq} \Delta V - \lambda^2 V = -\delta(\bx - \bxi) \,, \qquad V \to 0 \quad \mbox{as} \quad |\bx| \to \infty \,;\\[5pt]
 \label{V} V(\bx;\bxi,\lambda^2) \equiv \frac{1}{2\pi}K_0\left(\lambda |\bx - \bxi|\right) \,,
 \end{gather}
 \esub
 \noI where $K_0(z)$ is the modified Bessel function of the second kind. For $|z| = \mathcal{O}(1)$, the small and large argument asymptotics of $K_0(\lambda z)$ are
 \bsub
 	\begin{align} \label{Vsmall}	K_0(\lambda z) &\sim - \log\lambda - \log z + \log 2 - \gamma_e + \mathcal{O}(\lambda^2\log \lambda)\,, \qquad \lambda \to 0^+ \,,\\[5pt]
 \label{Vlarge} K_0(\lambda z) &\sim \sqrt{\frac{\pi}{2\lambda z}}e^{-\lambda z}\left(1 - \frac{1}{8}\frac{1}{\lambda z} + \frac{9}{128}\frac{1}{\lambda^2 z^2} + \mathcal{O}\left(\frac{1}{\lambda^3 z^3} \right)\right) \,, \qquad \lambda \to \infty \,,
 	\end{align}
 \esub
  	\noI where $\gamma_e \approx 0.5772$ is Euler's gamma constant. With \eqref{GRV} in \eqref{helmholtzGreens}, the regular problem for $\tilde{R}(\bx;\bxi,\lambda^2)$ is
 \BE \label{Req}
 	\Delta \tilde{R} - \lambda^2 \tilde{R} = 0 \,, \quad \bx \in \Omega \,; \qquad \partial_n \tilde{R} = - \partial_n V(\bx; \bxi, \lambda^2) \,, \quad x \in \partial \Omega \,.
 \EE
For general $\lambda^2$ and $\Omega$, \eqref{Req} can be solved numerically using a finite element method. When $\lambda^2$ is large, an asymptotic solution of \eqref{Req} will be developed (cf.~\S \ref{smallt}) to describe the short time behavior of capture time distribution. Once  $\tilde{R}$ is known, then \eqref{GRV} is used to calculate 
\bsub
\BE \label{Gjk}
	G_{j,k} = \frac{1}{2\pi} K_0\left(\lambda|\bx_j - \bx_k|\right) + \tilde{R}(\bx_j; \bx_k, \lambda^2) \,, \qquad j \neq k \,,
\EE
\noI for $G_{j,k}$ in \eqref{AkRk}. For the self-interaction term $R_{k,k}$ in \eqref{AkRk}, we use the local behavior of $G_h$ near $\bxi$ \eqref{helmholtzGreens_b} and the small argument asymptotic form of $K_0$ in \eqref{Vsmall} together with \eqref{GRV} to find
\BE \label{Rkk}
	R_{k,k} = \tilde{R}(\bx_k; \bx_k, \lambda^2) - \frac{1}{2\pi}\left(\log \lambda - \log 2 + \gamma_e\right) \,.
\EE
\esub
We remark that this asymptotic formulation in terms of Laplace transforms and Green's functions replaces a time-dependent problem \eqref{MainFreeProb}, having a singular initial condition, and sharp boundary layers on a non simply connected domain, with a time-independent and smooth problem \eqref{Req} posed on a hole free domain coupled with an inverse Laplace transform operation. The difficulty in this formulation arises when $s = \lambda^2$ in \eqref{Req} is large, or equivalently, when $t$ in \eqref{freedens} and \eqref{freedens} is small. In this case, the boundary data in \eqref{Req} becomes exponentially small in $\lambda$ according to the large argument asymptotics of $V$ in \eqref{Vlarge}. The solution $\tilde{R}$ then becomes obscured by the error associated with the numerical solver. An asymptotic method is thus needed to estimate $G_h$ in the large $\lambda^2$ limit.

Within this small time limit, the decomposition \eqref{GRV} suggests two different regimes of the free probability density. The first regime is captured by the free-space Green's function $V(\bx;\bx_0,\lambda^2)$, and sees only the exponentially few particles whose paths go directly from $\bx_0$ to $\partial \Omega_\eps$. The second regime, which requires the boundary contribution $\tilde{R}(\bx;\bx_0,\lambda^2)$, accounts for particles leaving whose paths have been influenced by $\partial \Omega$. In the next section, we compute a two-term asymptotic expansion of the solution to \eqref{Req} for $s = \lambda^2 \gg 1$, allowing for an accurate estimate of $P(t)$ and $C(t)$ for small time regimes. We also show that this estimate for small $t$, together with the numerical solution of \eqref{Req} for $s = \lambda^2$ and $t$ both $\mathcal{O}(1)$, can produce a uniformly valid estimate of $P(t)$ and $C(t)$ for all $t$. With the exception of one example in which $\Omega$ is an ellipse, for simplicity, we demonstrate this method for the case where $\Omega$ is the unit disk.
 
\begin{exmp} \label{ex1}
 In this example we verify the hybrid asymptotic method against a closed form separable solution of \eqref{MainFreeProb} on the unit disk. Let $\Omega$ be the unit disk  $\{ |\bx| \leq 1\}$ with a single circular trap of radius $\eps = 0.01$ centered at the origin and the walker initially at $\bx_0 = (0.3,0)$. The schematic and result are shown in Fig.~\ref{fig:ex1}. In Fig.~\ref{ex1fig}, the exact distribution from the separable solution is shown in solid. The small time estimate obtained from the hybrid method is shown in dashed (see \S \ref{smallt}).
\end{exmp}

\begin{figure}[htbp]
\centering
    \subfigure[Example \ref{ex1} schematic]{\label{ex1schem} \includegraphics[width=0.45\textwidth]{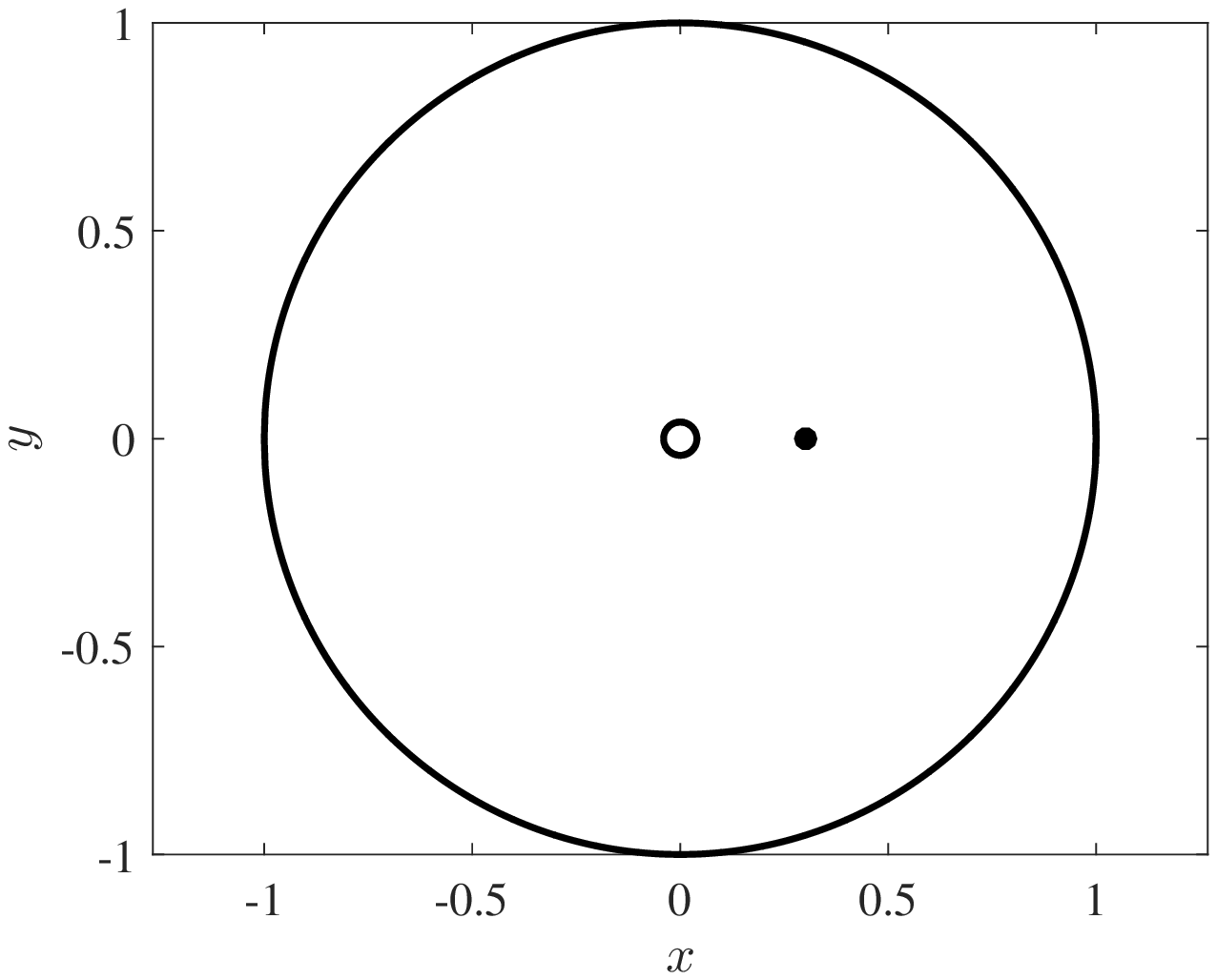}}
    \hspace{0.5cm}  
    \subfigure[$C(t)$]{\label{ex1fig}\includegraphics[width=0.45\textwidth]{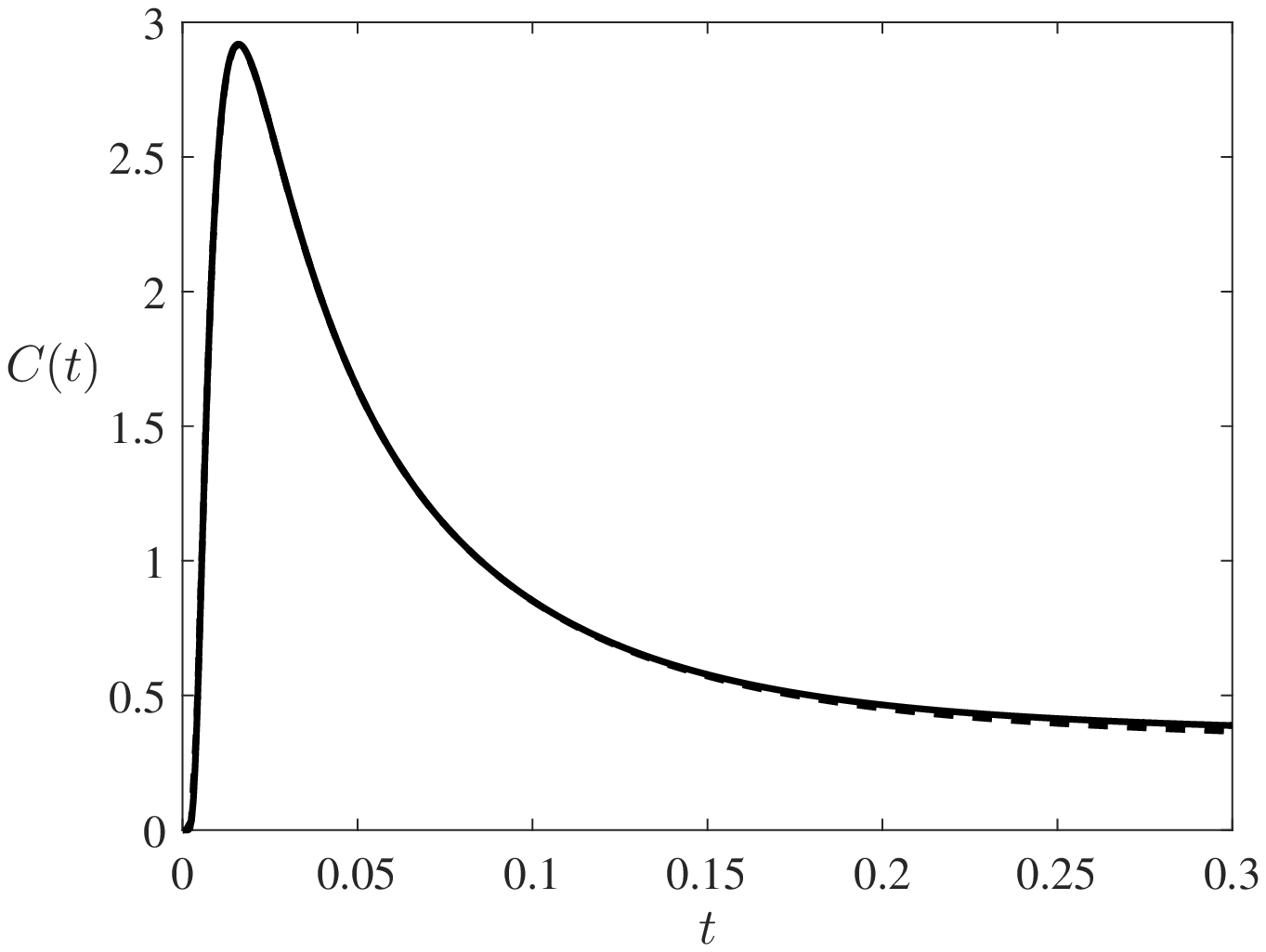}}
    \parbox{0.75\textwidth}{\caption{Schematic and result for Example \ref{ex1}. In (a), we show the schematic with one trap of radius $\varepsilon = 0.01$ centered at the origin (open circle), while the initial starting location for the particle is marked with a solid dot at $\bx_0 = (0.3, 0)$. In (b), we show the exact capture time distribution computed from an eigenfunction solution of \eqref{MainFreeProb} (solid). A small time estimate obtained from the hybrid method is shown in dashed. Very close agreement is observed when $t$ is small. \label{fig:ex1}}}
\end{figure}

\setcounter{equation}{0}
\section{Asymptotic estimate for $0 < t \ll 1$} \label{smallt}

We first analyze the time interval over which the particles that leave the domain through $\partial \Omega_\eps$ are dominated by those whose paths from $\bx_0$ to the closest point on $\partial \Omega$ are much shorter than those whose paths first hit $\partial \Omega$ before reaching any part of $\partial \Omega_\eps$. That is, we approximate \eqref{MainFreeProb} with the same problem on $\mathbb{R}^2$. In the Laplace domain, we analyze \eqref{AkRk} for $s = \lambda^2 \gg 1$ in \eqref{helmholtzGreens_a} assuming that the boundary contribution $\tilde{R}$ in \eqref{Gjk} is exponentially small in comparison to $V$. We thus have for $\lambda^2 \gg 1$, 

\BE \label{Gjkasymp}
	G_{j,k} \sim \frac{1}{2\pi} K_0\left(\lambda|\bx_j - \bx_k|\right)\,, \quad j \neq k \,; \qquad \lambda^2 \gg 1 \,.
\EE

\noI We verify this assumption below when we estimate $\tilde{R}$ using a boundary integral method. We also discard the exponentially small $\tilde{R}$ term in \eqref{Rkk} so that

\BE \label{Rkkasymp}
	R_{k,k} \sim - \frac{1}{2\pi}\left(\log \lambda - \log 2 + \gamma_e\right) \,. 
\EE

\noI From \eqref{Gjkasymp}, \eqref{Rkkasymp} and the large argument asymptotics of $K_0(\lambda z)$ in \eqref{Vlarge}, we observe that the off-diagonal elements of the Green's interaction matrix $\mathcal{G}$ are exponentially small in comparison to the diagonal elements. To leading order, we thus have

\BE \label{Akasymp}
	A_k = \frac{G_{k,0}}{1 + 2\pi\nu_k R_{k,k}}  \,, \qquad \lambda^2 \gg 1 \,,
\EE

\noI where $R_{k,k}$ is given asymptotically by \eqref{Rkkasymp}. To simplify the right-hand side of \eqref{Akasymp}, we denote $M$ as the number of traps whose centers $\bx_j$ satisfy $|\bx_0 - \bx_j| = \ell$, where $\ell$ is the distance from the starting location $\bx_0$ to the center(s) of the nearest trap(s)

\BE \label{ell}
	\ell \equiv \min_j |\bx_0 - \bx_j| \,; \qquad G_\ell \equiv \frac{1}{2\pi} K_0\left(\lambda\ell\right) \,.
\EE

\noI Then for all $j$ for which $|\bx_0 - \bx_j| > \ell$, we have that $G_{j,0}$ is exponentially smaller than $G_\ell$, where $G_\ell$ is defined in \eqref{ell}. Finally, assuming the traps are identical so that $\nu_j = \nu$ for each $j=1,\ldots,N$ and using (\ref{Gjkasymp} - \ref{ell}) in \eqref{LP}, we obtain the large $s$ (small time) asymptotics for $\mathcal{L}\lbrack P\rbrack(s)$ 

\BE \label{LPasymp}
	U(\bx_0,s) = \frac{1}{s} - \frac{M\nu}{1 - \nu\left(\frac{1}{2}\log s - \log 2 + \gamma_e\right)} \frac{K_0(\sqrt{s} \ell)}{s} \,.
\EE

\noI To compute $P(t)$, we perform a numerical inverse Laplace transform of $U(\bx_0,s)$ on \eqref{LPasymp}, which depends only on the distance $\ell$ from $\bx_0$ to the nearest trap(s), and the number of nearest traps $M$. It is independent of the shape of the domain and also the trap locations. The range of validity of \eqref{LPasymp}, however, may shrink for configurations in which the boundary plays a significant role along the paths of particles that first reach $\partial \Omega_\eps$. Typical scenarios include when $\bx_0$ is close to the boundary, the trap nearest $\bx_0$ is close to the boundary, or if $\bx_0$ and the nearest trap are separated by a bottleneck. In all of these cases, a significant portion of the particles that first reach $\partial \Omega_\eps$ have interacted with the boundary. In these cases, the infinite space problem is a poor approximation to \eqref{MainFreeProb}.

\begin{exmp} \label{boundaryfreedisk} Here we consider the case where $\Omega$ is a unit disk with initial location $\bx_0 = (0.2, 0)$ and five circular traps of radius $\varepsilon = 0.01$ centered at $\bx_j = \bx_0 + r_j (\cos\theta_j, \sin\theta_j)$ where 
\BE \label{traploc} r_{1,2,3} = 0.4\,, \quad (\theta_1, \theta_2, \theta_3) = \left(\frac{\pi}{6}, \frac{\pi}{2}, \pi\right) \,,\quad  (r_4, r_5) = (0.6, 0.8) \,, \quad (\theta_4, \theta_5) = \left(\frac{3\pi}{2}, \frac{5\pi}{4}\right)\,.
\EE 
\end{exmp}
The schematic of this configuration is shown in Fig. \ref{circschem}, where the starting location $\bx_0$ is marked by a solid dot, while the nearest (farther) traps are indicated by heavy (light) solid circles. In this case, we have $M = 3$ and $\ell = 0.4$. The asymptotic result obtained from \eqref{LPasymp} is plotted in heavy dashed in Fig. \ref{fig:smalltime}. The numerical result obtained from solving \eqref{MainFreeProb} with $\Omega$ being the unit disk and $\partial \Omega_\eps$ given by the trap locations above is plotted in heavy solid. Excellent agreement is observed.

\begin{figure}[htbp]
\centering
    \subfigure[Example \ref{boundaryfreedisk} schematic]{\label{circschem} \includegraphics[width=.45\textwidth]{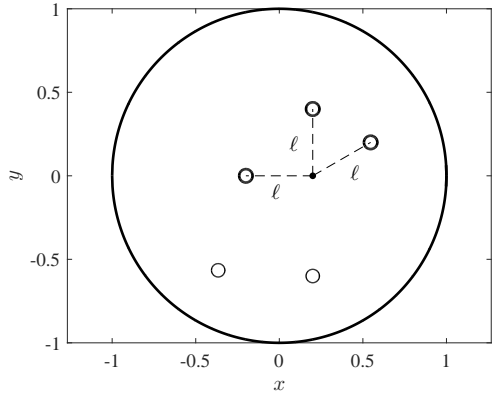}}
    \hspace{0.5cm}  
    \subfigure[Example \ref{boundaryfreesq} schematic]{\label{sqschem}\includegraphics[width=.45\textwidth]{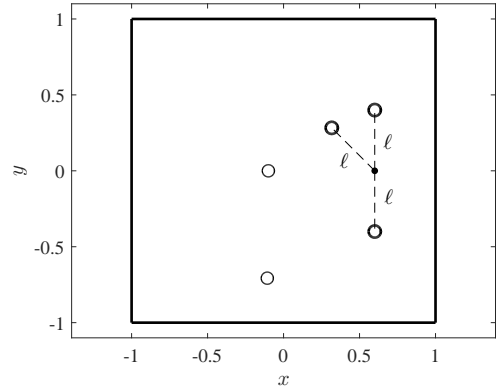}}
    \parbox{0.75\textwidth}{\caption{Schematics of configurations for Examples \ref{boundaryfreedisk} and \ref{boundaryfreesq}. In each figure, the starting locations $\bx_0$ for the random walk is marked by a solid dot, while the nearest (farther) traps are indicated by heavy (light) solid circles. In both cases, there are $M = 3$ nearest traps at a distance $\ell = 0.4$ from $\bx_0$. The traps have been enlarged for clarity.\label{fig:schem}}}
\end{figure}

\begin{exmp} \label{boundaryfreesq} Here we consider the case where $\Omega$ the square $[-1,1]^2$. The initial walker location is $\bx_0 = (0.6, 0)$ while same-sized traps are located at $\bx_j = \bx_0 + r_j (\cos\theta_j, \sin\theta_j)$, where $r_{1,2,3} = 0.4$, $(\theta_1, \theta_2, \theta_3) = (\pi/2, 3\pi/4, 3\pi/2)$, $(r_4, r_5) = (0.7, 1)$ and $(\theta_4, \theta_5) = (\pi, 5\pi/4)$. 
\end{exmp}
This configuration is shown in Fig.~\ref{sqschem}. As in Example \ref{boundaryfreedisk}, we have $M = 3$ and $\ell = 0.4$. The numerical result, given by \eqref{MainFreeProb} with $\Omega$ and $\partial \Omega_\eps$ as specified in Example \ref{boundaryfreesq}, is plotted in heavy dotted in Fig.~\ref{fig:smalltime}. As discussed above, the shape and size of the domain have a very small effect on the capture time distribution in the small time interval considered. As such, the boundary-free estimate given by \eqref{LPasymp} is valid for both the circular and square domains.

\begin{figure}[htbp]
\centering
    \subfigure[$P(t)$ for $t\ll 1$]{\label{smalltime_P} \includegraphics[width=.45\textwidth]{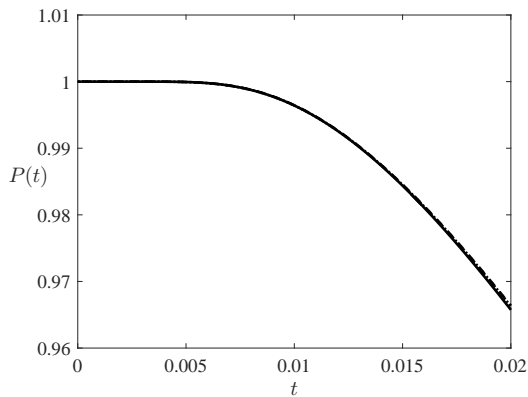}}  
    \hspace{0.5cm} 
    \subfigure[$C(t)$ for $t\ll 1$]{\label{smalltime_C}\includegraphics[width=.45\textwidth]{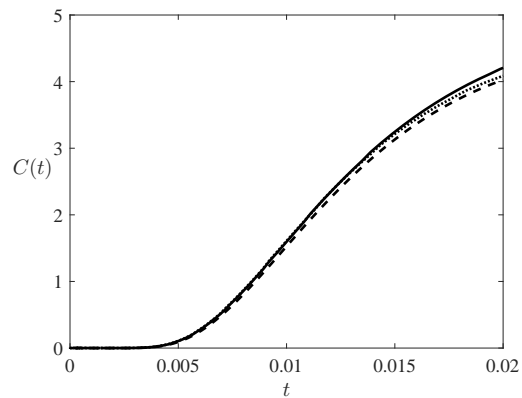}}
    \parbox{0.75\textwidth}{\caption{Results for Examples \ref{boundaryfreedisk} and \ref{boundaryfreesq}. In (a), we show $P(t)$ for $M = 3$, $\ell = 0.4$ and $\varepsilon = 0.01$ as numerically computed on the unit disk (solid) and square of side length 2 (dotted). The asymptotic estimate obtained from numerically inverting \eqref{LPasymp}, is plotted in dashed. The three lines are almost indistinguishable. Details given in the text. In (b), we show the corresponding plot for $C(t)$. As expected, the shape and size of the domain has little effect over the small time interval considered so that \eqref{LPasymp} provides a good estimate for both the circular and square domains. \label{fig:smalltime}}}
\end{figure}

We remark, however, that for the same $M$ and $\ell$, configurations for which the boundary effects are strong at small $t$ (e.g., when the initial location $\bx_0$ is close to $\partial \Omega$), the time interval over which the boundary-free estimate \eqref{LPasymp} is valid becomes much smaller. To capture boundary effects, thereby extending the range of validity of the asymptotic estimate for small time, we require an accurate estimate of $\tilde{R}$ for large $s = \lambda^2$. 

To do so, we employ a higher order hybrid numerical-asymptotic variation of the approach of \cite{kolokolnikov2003reduced}. The method is valid for a general domain with a smooth boundary. Our aim is to obtain an estimate of $\tilde{R}(\bx_j;\bx_k,s)$ for $G_{j,k}$ in \eqref{Gjk} in the limit of large $s = \lambda^2$, which accounts for particles that reflect off the boundary before reaching the $j$-th trap. We do not require the self-interaction term $\tilde{R}(\bx_k;\bx_k,s)$, as this term is small in comparison to the asymptotically large $\log s$ term in \eqref{Rkk}. We contrast our approach of asymptotically calculating the Helmholtz Green's function for large $s$ with that of \cite{godec2016first}, where the short time behavior of $C(t)$ is obtained by extracting the large $s$ asymptotics of an exact solution for $\mathcal{L}\lbrack C \rbrack (s)$.

We begin by setting $\bxi = \bx_0$ in \eqref{Req} then multiply by $V(\bx;\bx_j,s)$ and integrate over $\Omega$. Applying Green's identity and using \eqref{Veq} yields that
\BE \label{Rxj}
	\tilde{R}(\bx_j;\bx_0) = \int_{\partial\Omega} \! V(\bx^\prime;\bx_j)\partial_n\tilde{R}(\bx^\prime; \bx_0) - \tilde{R}(\bx^\prime; \bx_0)\partial_n V(\bx^\prime;\bx_j) \, d\bx^\prime \,,
\EE
\noI where we have dropped the $s$ dependence in the notation. All terms in \eqref{Rxj} are known, with $\partial_n\tilde{R}(\bx^\prime; \bx_0)$ specified by the boundary condition in \eqref{Req}, except for $\tilde{R}(\bx^\prime; \bx_0)$. It thus remains to estimate $\tilde{R}$. From \eqref{Rxj}, we observe that $\tilde{R}(\bx_j;\bx_0) \sim \mathcal{O}(e^{-\sqrt{s}\,(d_1 + d_2)})$, where $d_1$ is the distance from $\bx_0$ to the boundary, and $d_2$ is the distance from $\bx_j$ to the boundary. Therefore, if the shortest path from $\bx_0$ to $\bx_j$ is shorter than $d_1 + d_2$, we may discard $\tilde{R}$ in \eqref{GRV} in the short time estimates of $P(t)$ and $C(t)$. This was the basis for the boundary-free estimate \eqref{LPasymp}.

To estimate $\tilde{R}$, we observe that in the outer region away from the boundary, $\tilde{R} \sim 0$ to all orders in $\lambda$. For $s$ large, it therefore suffices to estimate $\tilde{R}$ in an $\mathcal{O}(\lambda^{-1})$ boundary layer near $\partial \Omega$. For a general domain, in terms of a local orthogonal coordinate system near the boundary where $\hat{\eta} = \lambda^{-1}\eta$ denotes the distance from $\bx$ to the boundary and $\xi$ the arc length along $\partial \Omega$, \eqref{Req} transforms to the following inner problem for $\tilde{R}$
\BE \label{laplbeltr}
	\lambda^2\partial_{\eta\eta} \tilde{R} - \lambda \frac{\kappa}{1-\kappa/\lambda}\partial_\eta \tilde{R} + \frac{1}{1-\kappa\eta/\lambda}\partial_\xi\left(\frac{1}{1-\kappa\eta/\lambda}\partial_\xi \tilde{R} \right) - \lambda^2\tilde{R} = 0\,.
\EE
\noI In \eqref{laplbeltr}, $\kappa = \kappa(\xi)$ is the curvature of $\partial \Omega$. When $\Omega$ is the unit disk, \eqref{laplbeltr} simplifies to the Laplacian in polar coordinates 
\BE \label{Reqpol}
	\lambda^2\partial_{\eta\eta} \tilde{R} - \lambda \frac{1}{1-\eta/\lambda}\partial_\eta \tilde{R} + \frac{1}{(1-\eta/\lambda)^2} \partial_{\xi\xi} \tilde{R} - \lambda^2\tilde{R} = 0\,.
\EE
\noI While we demonstrate the method on \eqref{Reqpol} mainly for the unit disk, the same technique may be used to estimate $\tilde{R}$ for \eqref{laplbeltr} on an arbitrary domain (see \eqref{a1b1gen} and Example \ref{ellipse}). To begin, we first expand \eqref{Reqpol} for large $\lambda$ and use $\partial_n = -\lambda\partial_\eta$ in the boundary condition of \eqref{Req} to obtain
\bsub \label{Reqasymp}
\begin{gather} 
\label{Reqasympeq} \lambda^2\partial_{\eta\eta}\tilde{R} - \lambda\left(1 + \frac{\eta}{\lambda} + \frac{\eta^2}{\lambda^2}\right)\partial_\eta\tilde{R} + \left(1 + 2\frac{\eta}{\lambda} + 3\frac{\eta^2}{\lambda^2}\right)\partial_{\xi\xi}\tilde{R} - \lambda^2\tilde{R} + \mathcal{O}(\lambda^{-3}) = 0\,; \\[5pt]
\label{Reqasympbc} 
\partial_\eta\tilde{R} \left|_{\eta = 0} \right. = \frac{e^{-\lambda|\br|}}{2\sqrt{2\pi\lambda|\br|}}\left\lbrack - \left\langle \frac{\br}{|\br|},\hat{\bn} \right\rangle -\frac{3}{8\lambda} \left\langle \frac{\br}{|\br|^2},\hat{\bn} \right\rangle + \mathcal{O}(\lambda^{-2}) \right\rbrack \,,
\end{gather}
\esub
\noI where $\hat{\bn}$ denotes the outward unit normal on $\partial \Omega$, $\br = \bx^\prime - \bx_0$, and $\left\langle \cdot,\cdot\right\rangle$ denotes the dot product. In \eqref{Reqasympbc}, we have used the large argument asymptotic forms for $K_0(z)$ and $K_1(z)$, while $\bx^\prime$ is parameterized by the arc length $\xi$ and restricted to the boundary. According to \eqref{Reqasymp}, we expand $\tilde{R}$ as
\BE \label{Rexp}
	\tilde{R} \sim \tilde{R_0} + \frac{1}{\lambda}\tilde{R}_1 + \cdots\,.
\EE
\noI With \eqref{Rexp} in \eqref{Reqasympbc} and matching orders of $\lambda$, we obtain
\bsub\label{BCexp}
\BE \label{BCexp1}
	\partial_\eta\tilde{R}_0 \left|_{\eta = 0} \right. = -\left\langle \frac{\br}{|\br|},\hat{\bn} \right\rangle \frac{e^{-\lambda|\br|}}{2\sqrt{2\pi\lambda|\br|}} \,,
\EE
\BE \label{BCexp2}
	\partial_\eta\tilde{R}_1 \left|_{\eta = 0} \right. = -\frac{3}{8}\left\langle \frac{\br}{|\br|^2},\hat{\bn} \right\rangle \frac{e^{-\lambda|\br|}}{2\sqrt{2\pi\lambda|\br|}} \,.
\EE
\esub
For $\tilde{R}_0$, we account for the $\mathcal{O}(\lambda)$ variation of the boundary data by making an ansatz of the form
\bsub
\BE \label{R0ans}
	\tilde{R}_0 = f_0(\eta, \xi)e^{-\lambda|\br|}	\,; \qquad f_0(\eta,\xi) = c_0(\xi) S_0(\eta, \xi) \,.
\EE
\noI Comparing \eqref{R0ans} with \eqref{BCexp1}, we set
\BE \label{c0}
	c_0(\xi) = -\left\langle \frac{\br}{|\br|},\hat{\bn} \right\rangle\frac{1}{2\sqrt{2\pi\lambda|\br|}} \,; \qquad \partial_\eta S_0 \left|_{\eta = 0} \right. = 1 \,.
\EE
\esub
\noI Substituting \eqref{R0ans} into \eqref{Reqasymp}, canceling the common exponential prefactors and collecting $\mathcal{O}(\lambda^2)$ terms, we obtain the ODE for $S_0(\eta)$
\begin{subequations}
\BE \label{S0eq}
\partial_{\eta\eta}S_0 - \phi(\xi)^2 S_0 = 0 \,, \qquad \partial_\eta S_0 \left|_{\eta = 0} \right. = 1 \,, \quad S_0 \to 0 \enspace \mbox{as} \enspace \eta \to \infty\,,
\EE
\noI where we have defined
\BE \label{phi}
	 \phi(\xi) \equiv \sqrt{1-|\br|_\xi^2} \,; \qquad |\br|_\xi < 1 \,.
\EE
\end{subequations}
\noI The limiting condition in \eqref{S0eq} is required for matching to the outer solution $\tilde{R} \sim 0$. Note that the $|\br|_\xi^2$ term in \eqref{S0eq} arises from the $\partial_{\xi\xi}$ term in \eqref{Reqasympeq}, which becomes $\mathcal{O}(\lambda^2)$ due to the fast variation of the boundary data in \eqref{Reqasympbc}. The solution of \eqref{S0eq} is
\BE \label{S0}
	S_0(\eta,\xi) = B_0(\xi)e^{-\phi \eta} \,; \qquad B_0(\xi) = -\frac{1}{\phi(\xi)} \,.
\EE
\noI We therefore have that the solution for $\tilde{R}_0$ near the boundary is given by \eqref{R0ans} with $c_0(\xi)$ and $S_0(\eta,\xi)$ defined in \eqref{c0} and \eqref{S0}, respectively. The leading order behavior of $\tilde{R}(\bx^\prime;\bx_0)$ on the boundary in \eqref{Rxj} is then given by $\tilde{R}(\bx^\prime,\bx_0) \sim c_0(\xi) B_0(\xi)$.
For $\tilde{R}_1$, we make the ansatz
\bsub
\BE \label{R1ans}
	\tilde{R}_1 = f_1(\eta, \xi)e^{-\lambda|\br|}	\,; \qquad f_1(\eta,\xi) = c_1(\xi) S_1(\eta, \xi) \,,
\EE
\noI and set
\BE \label{c1}
	c_1(\xi) = -\frac{3}{8}\left\langle \frac{\br}{|\br|^2},\hat{\bn} \right\rangle\frac{1}{2\sqrt{2\pi\lambda|\br|}} \,; \qquad \partial_\eta S_1 \left|_{\eta = 0} \right. = 1 \,.
\EE
\esub
\noI Substituting \eqref{R0ans} and \eqref{R1ans} into \eqref{Reqasymp} and collecting $\mathcal{O}(\lambda)$ terms, we obtain for $S_1$
\bsub
\BE \label{S1eq}
\partial_{\eta\eta}S_1 - \phi(\xi)^2 S_1 = \frac{c_0}{c_1}\left\lbrack a_1 + b_1\eta\right\rbrack S_0 \,,\qquad \partial_\eta S_1 \left|_{\eta = 0} \right. = 1 \,, \quad S_1 \to 0 \enspace \mbox{as} \enspace \eta \to \infty\,,
\EE
\noI where
\BE \label{a1b1}
	a_1 = -\phi + 2|\br|_\xi \frac{c_0^\prime}{c_0} - 2|\br|_\xi\frac{\phi^\prime}{\phi} + |\br|_{\xi\xi} \,, \qquad b_1 = -2|\br|_\xi^2-2|\br|_\xi\phi^\prime \,.
\EE
\esub
\noI We solve the ODE \eqref{S1eq} for $S_1$ to obtain
\BE \label{S1}
	S_1(\eta, \xi) = B_1(\xi) e^{-\phi\eta} + \frac{c_0\eta}{4\phi^3 c_1}\left\lbrack 2\phi a_1 + b_1 + \phi b_1\eta \right\rbrack e^{-\phi\eta} \,; \qquad B_1(\xi) = \frac{1}{\phi} \left\lbrack \frac{c_0\left(2\phi a_1 + b_1 \right)}{4\phi^3 c_1} -1\right\rbrack \,,
\EE
\noI where $a_1$ and $b_1$ are defined in \eqref{a1b1}.
%
%
%
%
%
%
%
%
%
%
We conclude that a two-term expansion for $\tilde{R}(\bx^\prime;\bx_0)$ accurate to $\mathcal{O}(\lambda^{-1}) \sim \mathcal{O}(s^{-1/2})$ on $\partial \Omega$ in \eqref{Rxj} is given by
\BE \label{Rbxp}
	\tilde{R}(\bx^\prime;\bx_0) \sim c_0(\xi)B_0(\xi) + \frac{1}{\lambda}c_1(\xi)B_1(\xi) \,,
\EE
\noI where $B_0$ and $B_1$ are given in \eqref{S0} and \eqref{S1}, respectively. While the integral in \eqref{Rxj} may be evaluated asymptotically using Laplace's method, we evaluate it numerically in order to preserve the $\mathcal{O}(\lambda^{-1})$ accuracy. 

We now compare estimates for $P(t)$ and $C(t)$ on the unit disk with one trap obtained from using the one- and two-term expansions for $\tilde{R}(\bx^\prime;\bx_0)$ on the boundary. For $N = 1$ in \eqref{AkRk} and $s \gg 1$, $A_1$ simplifies to
\bsub
\BE \label{A1}
	A_1(s) = \frac{ G_{1,0}}{1+2\pi\nu R_{1,1}} \,,
\EE
\noI where
\BE \label{G10}
	G_{1,0} = \frac{1}{2\pi}K_0\left(\sqrt{s}\,|\bx_0 - \bx_1|\right) + \tilde{R}(\bx_1;\bx_0,s) \,,
\EE
\noI and
\BE \label{R11}
 	R_{1,1} \sim -\frac{1}{2\pi}\left(\frac{1}{2}\log s - \log 2 + \gamma \right) \,.
\EE
\esub
\noI In \eqref{G10}, $\tilde{R}(\bx_1;\bx_0,s)$ is obtained from \eqref{Rxj} using either a one- or two-term expansion for $\tilde{R}(\bx^\prime;\bx_0)$ obtained above. In \eqref{R11}, we have dropped the exponentially small self-interaction term $\tilde{R}(\bx_1;\bx_1,s)$. The free probability $P(t)$ and capture time distribution $C(t)$ are then given by \eqref{LPC}. We illustrate the theory in the following example.

\begin{exmp} \label{onetrap} In this example, we consider a single trap with radius $\eps = 0.01$ and center $\bx_1 = r_1(\cos \theta_1, \sin\theta_1)$ with $r_1 = 0.6$ and $\theta_1 = 3\pi/4$. The Brownian particle starts at $\bx_0 = (0.7, 0)$. Results shown in Fig.~\ref{fig:twoterm}.
\end{exmp}
The numerical result is shown in heavy solid in Fig.~\ref{fig:twoterm}. For such a configuration, and for time extending beyond the initial flux of particles that travel straight from $\bx_0$ to the trap, the boundary is expected to significantly impact the particles that get captured. This is evidenced by the inability of the boundary-free approximation (dotted) to predict the behavior of $P(t)$ and $C(t)$ beyond a small initial time interval. The one-term approximation (dash-dotted) slightly extends the range of validity of the estimate. The time at which the boundary-free and one-term estimates begin to diverge may be regarded as the approximate time at which particles that have interacted with the boundary first begin to become trapped. In contrast to the previous two estimates, the two-term approximation (dashed) accurately predicts $C(t)$ to almost $t \sim \mathcal{O}(1)$. In fact, it is able to predict up to and beyond the mode of $C(t)$, the most common capture time.

\begin{figure}[htbp]
\centering
    \subfigure[Example \ref{onetrap} schematic]{\label{onetrapschem} \includegraphics[width=.475\textwidth]{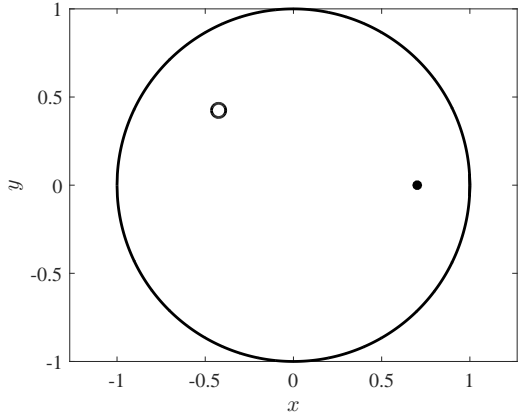}}  
    \hspace{0.5cm} 
    \subfigure[$C(t)$]{\label{twotermC}\includegraphics[width=.475\textwidth]{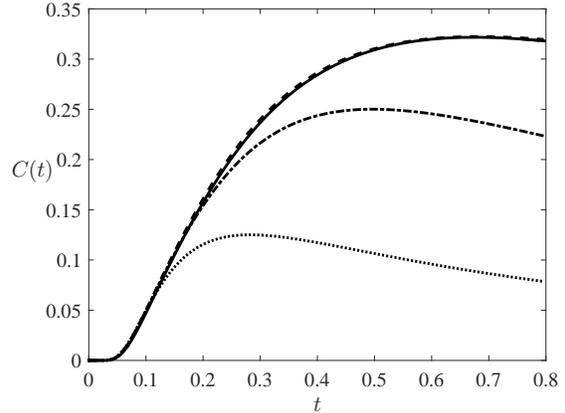}}
    \parbox{0.75\textwidth}{\caption{Schematic and results for Example \ref{onetrap}. In (a), we show the schematic with one trap of radius $\varepsilon = 0.01$ centered at $\bx_1 = r_1(\cos \theta_1, \sin\theta_1)$ with $r_1 = 0.6$ and $\theta_1 = 3\pi/4$ (open circle). The initial walker location (solid dot) is $\bx_0 = (0.7, 0)$. In (b), we show the numerically computed capture time density (solid) along with the boundary-free approximation (dotted), one-term (dash-dotted), and two-term estimates (dashed). The two-term estimate is almost indistinguishable from the numerical solution of \eqref{MainFreeProb}. Note that the two-term estimate is able to accurately predict the location of the mode of $C(t)$. \label{fig:twoterm}}}
\end{figure}

\begin{exmp} \label{multitrap} Here we demonstrate the method on the multi-trap scenario of Fig. \ref{fig:smalltime}. Again, we specify the starting location $\bx_0 = (0.2, 0)$ and trap locations $\bx_j = \bx_0 + r_j(\cos\theta_j,\sin\theta_j)$, where $r_j$ and $\theta_j$ are given in \eqref{traploc}. Results in Fig.~\ref{fig:multi}.
\end{exmp}
For example~\ref{multitrap} we solve the full system \eqref{AkRkmat} for $\{A_1(s), \ldots, A_5(s)\}$, with $G_{j,k}$ and $R_{k,k}$ given by \eqref{Gjk} and \eqref{Rkk}, respectively. In both \eqref{Gjk} and \eqref{Rkk}, $\tilde{R}$ is given in terms of the integral \eqref{Rxj} with its value on the boundary given by \eqref{Rbxp}. Using \eqref{LPC}, we apply a numerical inverse transform to compute $P(t)$ and $C(t)$. The asymptotic result is shown in Fig. \ref{fig:multi} (dashed) and is compared to that obtained from numerically solving \eqref{MainFreeProb} (solid), and from the boundary-free approximation (dotted) obtained from solving the system \eqref{AkRkmat} using \eqref{Gjkasymp} for $G_{i,j}$ and \eqref{Rkk} with $\tilde{R} = 0$ for $R_{k,k}$. We observe excellent agreement between the two-term asymptotic estimate and the numerical result. In fact, both the two-term estimate and the boundary-free approximation almost exactly capture the mode of $C(t)$. This may be attributed to the proximity of $\bx_0$ to  the nearest traps, and consequently, the limited contribution from the boundary. We remark, however, that the accuracy of the two-term asymptotic estimate persists even to only moderately small values of $t$, well beyond the range that the boundary-free approximation can capture.

\begin{figure}[htbp]
\centering
    \subfigure[$P(t)$]{\label{multiP} \includegraphics[width=.475\textwidth]{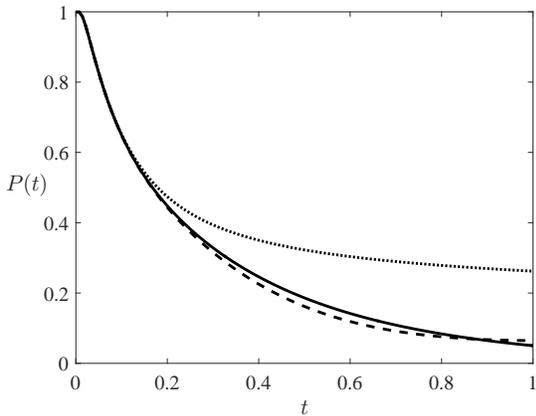}}  
    \hspace{0.5cm} 
    \subfigure[$C(t)$]{\label{multiC}\includegraphics[width=.475\textwidth]{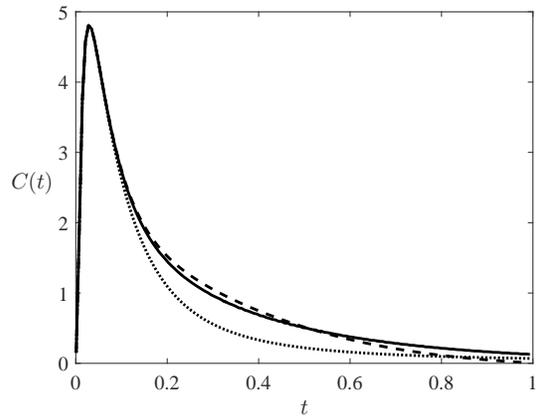}}
    \parbox{0.75\textwidth}{\caption{Results for Example \ref{multitrap}. In (a), we show the free probability density $P(t)$ for a particle on a unit disk starting a random walk from $\bx_0 = (0.2,0)$. The locations of the five circular traps of radius $\eps = 0.01$ are given by \eqref{traploc}. The full numerical result is shown in solid, the two-term asymptotic result in dashed, and boundary-free approximation in dotted. In (b), we show the corresponding plots for $C(t)$. While both the boundary-free and two-term estimates are able to capture the mode of $C(t)$, the agreement of the two-term estimate persists well past the mode and into the tails of $P(t)$ and $C(t)$. \label{fig:multi}}}
\end{figure}

\begin{exmp} \label{boundaryeffect} Here we illustrate that the two-term estimate is able to capture the strong boundary effects of when the starting location $\bx_0$ is near $\partial \Omega$. With the starting location at $\bx_0 = (0.92, 0)$, we set the locations of the five circular traps of radius $\eps = 0.01$ at $\bx_j = \bx_0 + r_j  (\cos\theta_j,\sin\theta_j)$ where
\BE \label{traploc_boundary} r_{1,2,3} = 0.4\,, \quad (\theta_1, \theta_2, \theta_3) = (3\pi/4, \pi, 5\pi/4) \,,\quad  (r_4, r_5) = (0.6, 0.8) \,, \quad (\theta_4, \theta_5) = (7\pi/6, 5\pi/6)\,.
\EE 
Results in Fig.~\ref{fig:multi_boundary}.
\end{exmp}
In Fig.~\ref{fig:multi_boundary}, we observe that the two-term estimate (dashed) more closely predicts $P(t)$ and $C(t)$ over a much longer time interval than does the boundary-free approximation (dotted). Compared to Fig.~\ref{fig:multi}, the boundary-free approximation is an especially poor predictor at moderate values of $t$ due to the presence of strong boundary effects. This implies that over the time interval in which the boundary-free approximation diverges from the numerical and two-term results in Fig.~\ref{fig:multi_boundary}, a significant proportion of absorbed particles had first interacted with the boundary. As a result, the boundary-free approximation over-estimates the survival probability, as it does not account for the ``funneling'' by the boundary of particles into the traps. Conversely, the two-term estimate predicts $P(t)$ and $C(t)$ for a longer time interval, and also reasonably predicts the mode of $C(t)$ (the time and the frequency). However, compared to Fig. \ref{fig:multi}, the interval of agreement is noticeably smaller. The reason for this discrepancy is that the expansion in the boundary condition \eqref{Reqasympbc} requires that $\sqrt{s}\,|\bx^\prime - \bx_0| \gg 1$. For $\bx^\prime \in \partial \Omega$ near $\bx_0$, where the contribution to the integral \eqref{Rxj} is greatest, the requirement for large $s$ satisfying $\sqrt{s}\,|\bx^\prime - \bx_0| \gg 1$ limits the range of validity to very small $t$. However, we note that a full numerical solution of \eqref{Req} in this case is less likely to be distorted by numerical error, since the maximum contribution of the boundary data is no longer exponentially small when $\sqrt{s}\,|\bx^\prime - \bx_0|$ falls outside the asymptotic regime. As such, it is still possible to combine the finite element method with the asymptotic solution to obtain a uniform estimate of $P(t)$ and $C(t)$.

\begin{figure}[htbp]
\centering
    \subfigure[Example \ref{boundaryeffect} schematic]{\label{multiP_boundary} \includegraphics[width=.45\textwidth]{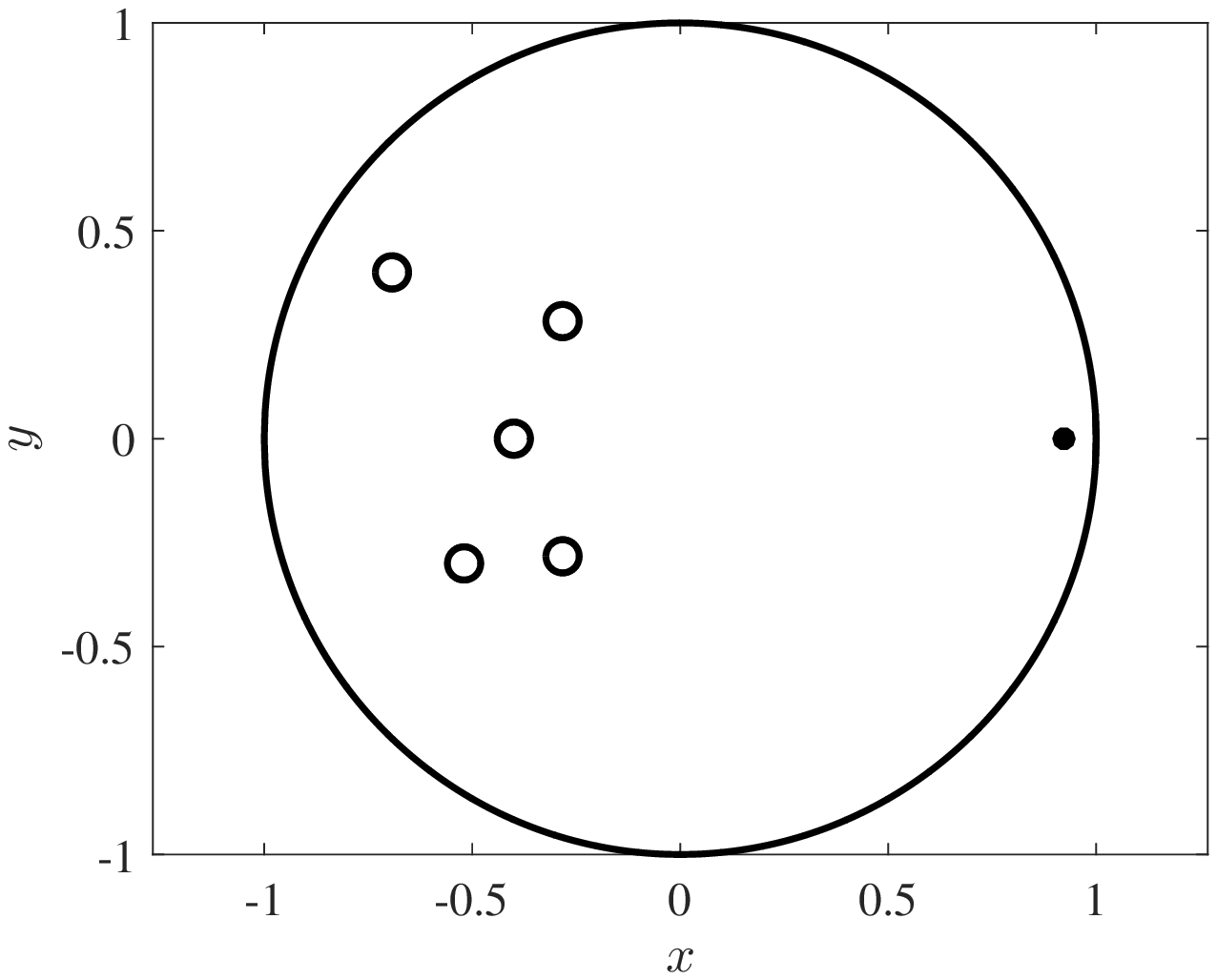}}  
    \hspace{0.5cm} 
    \subfigure[$C(t)$]{\label{multiC_boundary}\includegraphics[width=.45\textwidth]{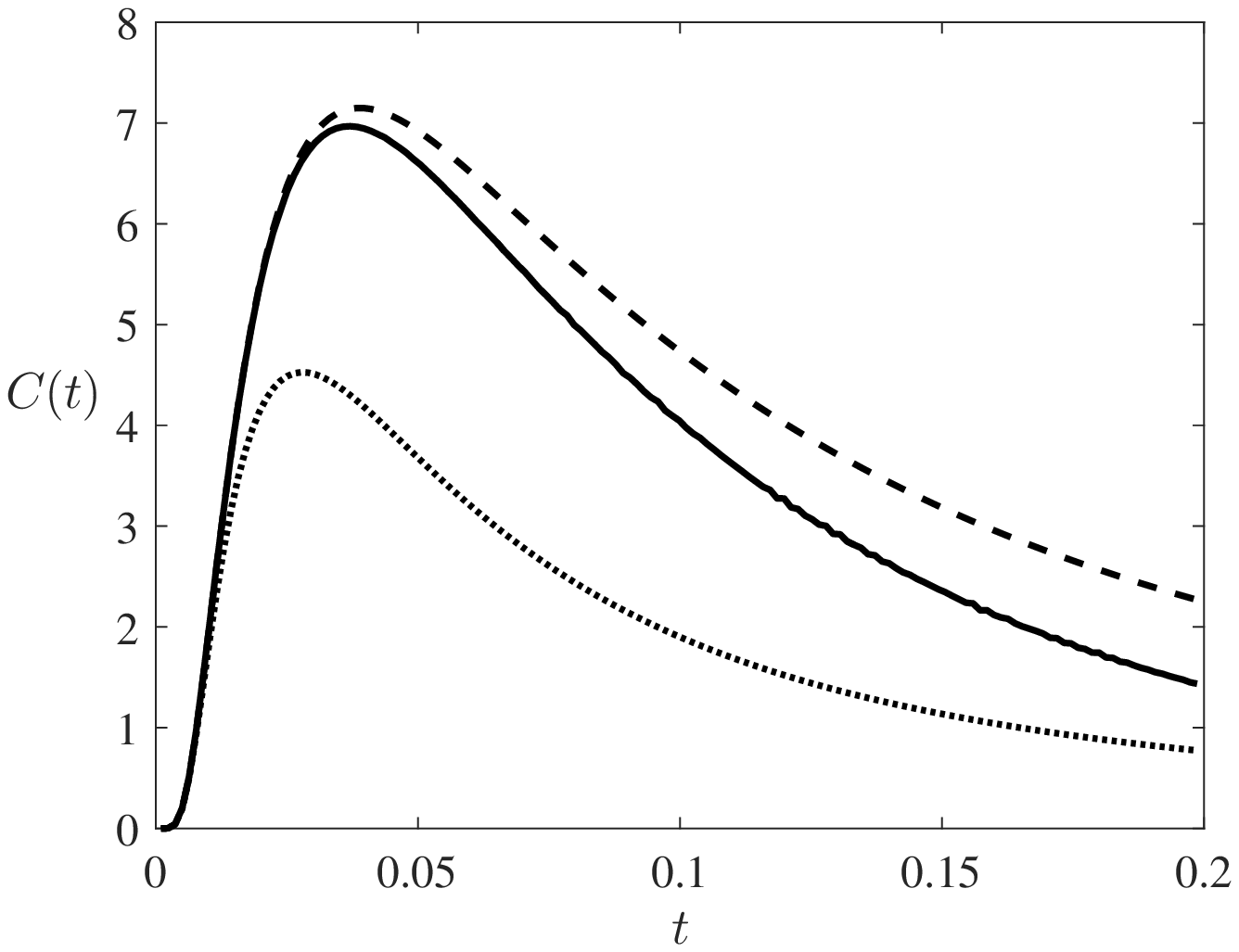}}
    \parbox{0.75\textwidth}{\caption{Schematic and results for Example \ref{boundaryeffect}. In (a), we show the schematic with $\bx_0 = (0.92,0)$ (solid dot) along with the locations of the five circular traps of radius $\eps = 0.01$ given by \eqref{traploc_boundary} (open circles). In (b) we show the full numerical result for $C(t)$ (solid), the two-term asymptotic estimate (dashed), and the boundary-free approximation (dotted). While the two-term estimate better predicts the full result over a longer time interval than does the boundary-free approximation, the interval of agreement is smaller than that seen in Fig.~\ref{fig:multi} due to the proximity of $\bx_0$ to the boundary.  \label{fig:multi_boundary}}}
\end{figure}

%

In the next two examples, we demonstrate that the large $s$ (small time) estimate for $G_h$ given by \eqref{GRV} with \eqref{Rbxp} is capable of capturing $C(t)$ when it is bimodal. In such distributions, the first peak in $C(t)$ is due to the first wave of particles being caught by a trap near the starting location $\bx_0$. The second peak is due to the next wave of particles being caught by the second absorbing set located opposite the first trap and at distance farther from $\bx_0$. One such configuration that produces a bimodal distribution is shown in Fig.~\ref{sixtrapschem}, where the starting location is indicated by the solid dot while the open circles represent equal sized traps. Critically, the second absorbing set must be larger than the first trap, casting a ``wider net'' to accommodate the increased dispersal of particles. One way to achieve this effect with a smaller number of traps is by replacing the traps of the second absorbing set with fewer larger traps (Fig.~\ref{fourtrapschem}).

\begin{exmp} \label{multimode1}
In this example, we set the starting location at $\bx_0 = (0,0)$, with the nearest trap centered at $\bx_1 = (0.3, 0)$. The five traps composing the second absorbing set are centered on the ring of radius $r_c = 0.71$ at angles $3\pi/4, 7\pi/8, \pi, 9\pi/8$, and $5\pi/4$. All traps share a common radius of $\eps_c = 0.01$. This configuration is shown in Fig.~\ref{sixtrapschem}, while the bimodal distribution it produces is shown in Fig.~\ref{multimodesamesize}.
\end{exmp}

\begin{figure}[htbp]
\centering
    \subfigure[equally sized traps]{\label{sixtrapschem} \includegraphics[width=.475\textwidth]{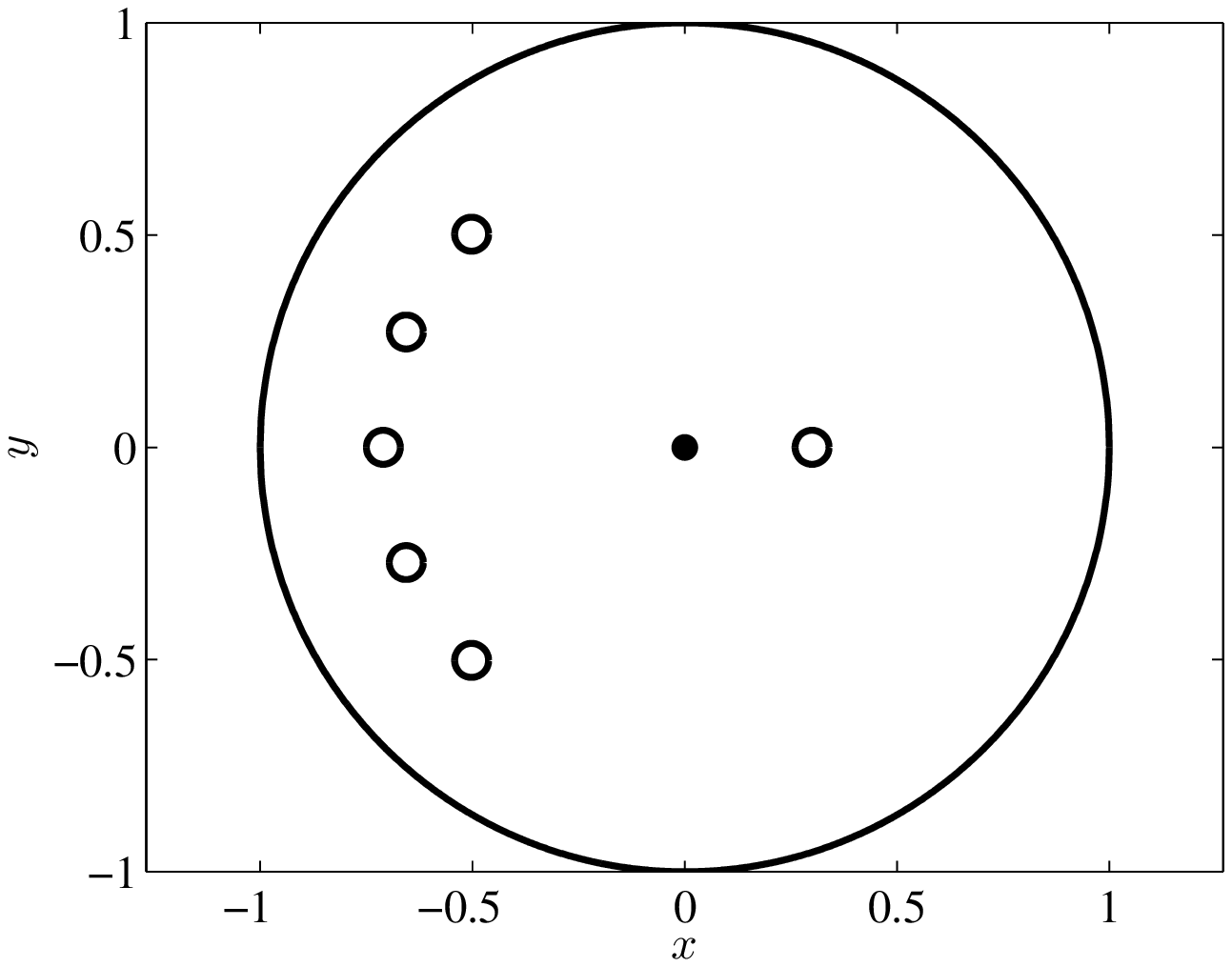}}  
    \hspace{0.5cm} 
    \subfigure[bimodal $C(t)$]{\label{multimodesamesize}\includegraphics[width=.475\textwidth]{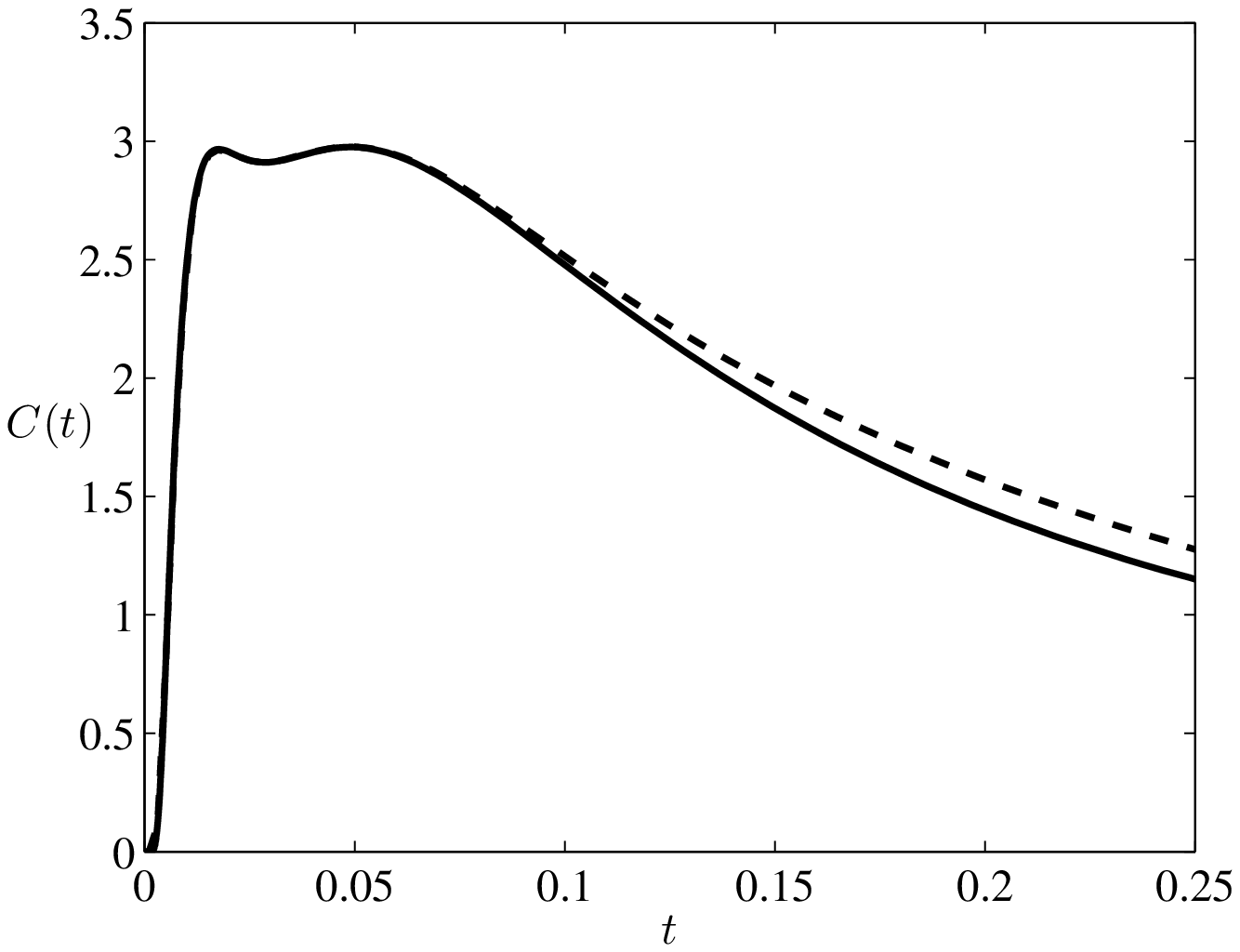}}
    \parbox{0.75\textwidth}{\caption{Schematic and results for Example \ref{multimode1}. In (a), we show the starting location $\bx_0$ at the origin (solid dot) surrounded by six traps of equal size that form two absorbing sets. The first set consists of the nearest trap centered at $(r_1,0)$. The second set is composed of the union of the five traps centered on a ring $r_c > r_1$. In (b), we show the resulting bimodal distribution for $C(t)$. The first peak corresponds to paths straight from $\bx_0$ to $(r_1,0)$. The second peak corresponds to direct paths from $\bx_0$ to the second absorbing set. The second absorbing set needs to be larger due to the greater dispersal of paths on the ring $r = r_c$ versus $r = r_1$. \label{fig:multimode}}}
\end{figure} 

In Fig.~\ref{fig:multimode_time}, we show snapshots of $p(\bx,t)$ near the two times corresponding to the peaks of $C(t)$ in Fig.~\ref{multimodesamesize}. The plots were generated from numerical solution of \eqref{MainFreeProb} with the initial location $\bx_0$ and absorbing set $\partial \Omega_\eps$ as specified in Example \ref{multimode1}. The first peak in Fig.~\ref{multimodesamesize} occurs when a significant portion of the distribution first reaches the first trap (Fig.~\ref{sixtrap_p_time_p02}). As the front traveling toward positive $x$-values spreads past the first trap, the frequency of particles exiting declines until the front on the opposite side spreads to the second absorbing set composed of the five traps in the second and third quadrants (Fig.~\ref{sixtrap_p_time_p05}). This leads to an increase in the rate of particles exiting the domain, generating the second peak in Fig.~\ref{multimodesamesize}. This is followed by a monotonic decay of $C(t)$ as the particle distribution becomes more diffuse and uniform.

\begin{figure}[htbp]
\centering
    \subfigure[$p(\bx, t)$ at $t= 0.02$.]{\label{sixtrap_p_time_p02} \includegraphics[width=.475\textwidth]{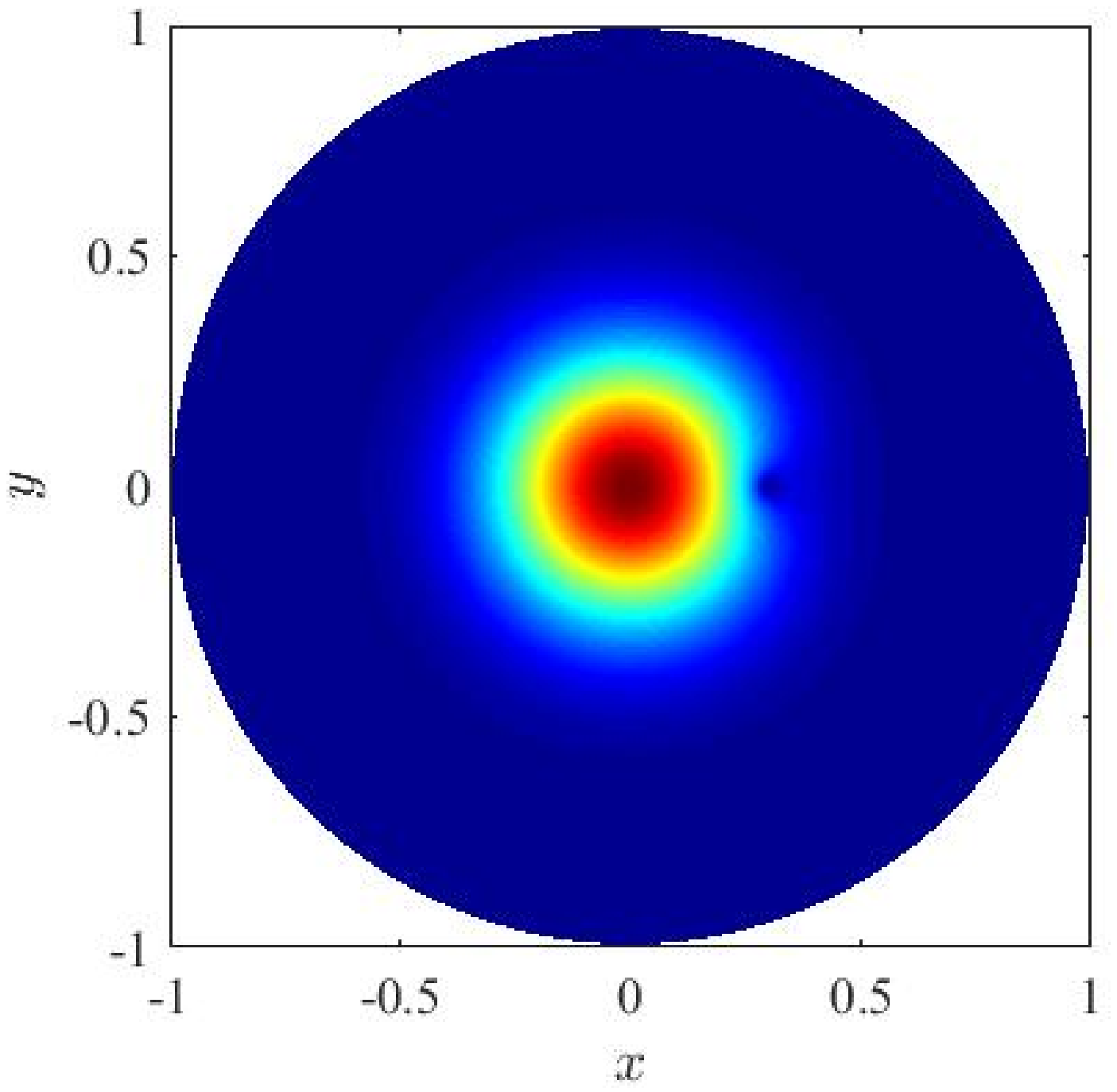}}  
    \hspace{0.5cm} 
    \subfigure[$p(\bx, t)$ at $t= 0.05$.]{\label{sixtrap_p_time_p05}\includegraphics[width=.475\textwidth]{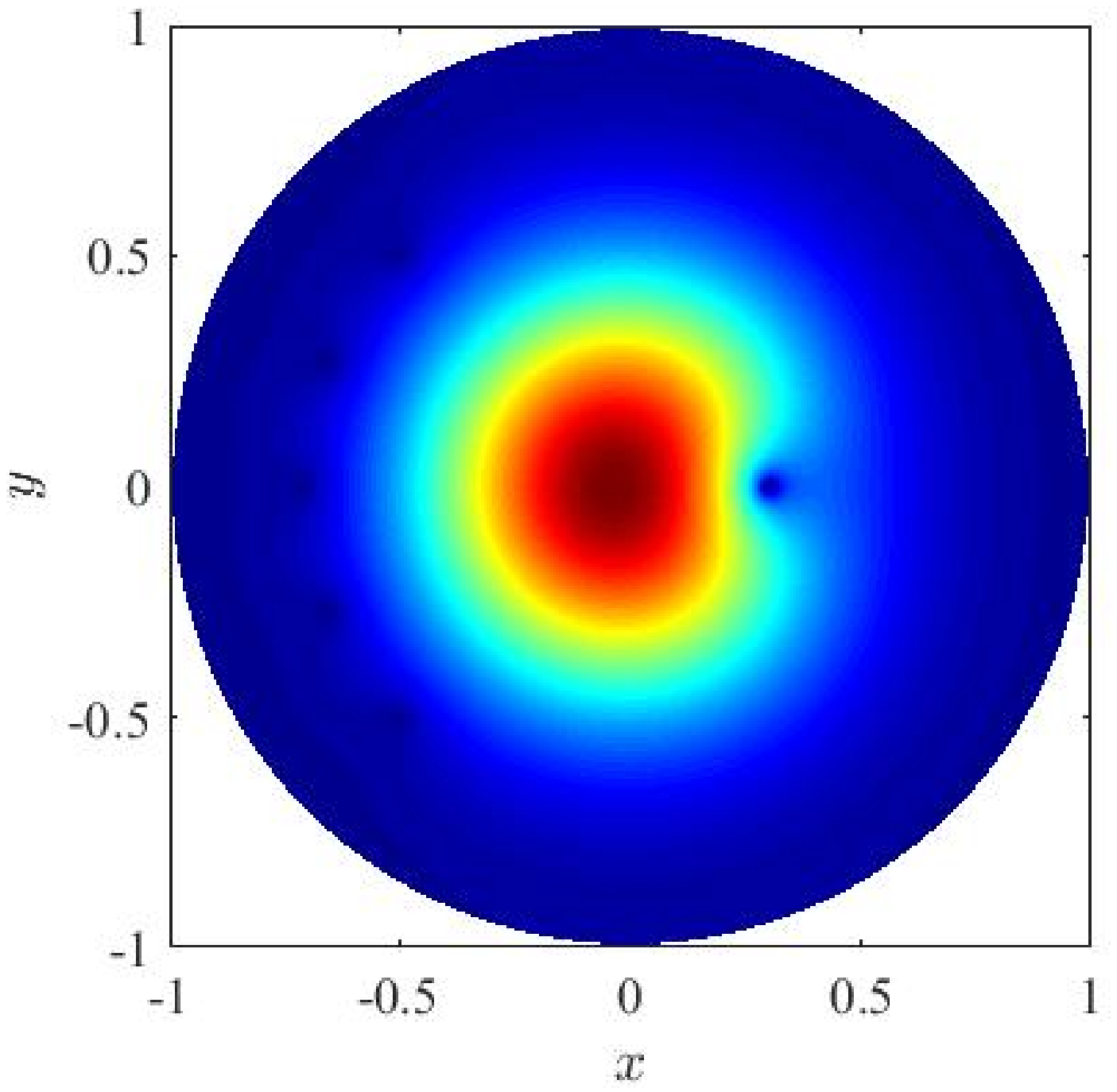}}
    \parbox{0.75\textwidth}{\caption{Snapshots of solution $p(\bx,t)$ of \eqref{MainFreeProb} at (a) $t = 0.02$ and (b) $t = 0.05$ for Example \ref{multimode1}. The times of the two snapshots approximately correspond to the times of the two peaks in $C(t)$ in Fig.~\ref{multimodesamesize}. The first peak occurs as the initial distribution reaches the nearest trap, while the second peak occurs later time when the distribution spreads to the second set of five absorbing traps. \label{fig:multimode_time}}}
\end{figure}

\begin{exmp} \label{multimode2}
In this example, we set the starting location at $\bx_0 = (0,0)$, with the nearest trap of radius $\eps_1 = 0.005$ centered at $\bx_1 = (0.3, 0)$. The three larger traps of common radius $\eps_c = 5\eps_1$ composing the second absorbing set are centered on the ring of radius $r_c = 0.75$ at angles $2\pi/3, \pi$, and $4\pi/3$. This configuration is shown in Fig.~\ref{fourtrapschem}, while the bimodal distribution it produces is shown in Fig.~\ref{multimodediffsize}.
\end{exmp}

\begin{figure}[htbp]
\centering
    \subfigure[Traps of differing radii.]{\label{fourtrapschem} \includegraphics[width=.475\textwidth]{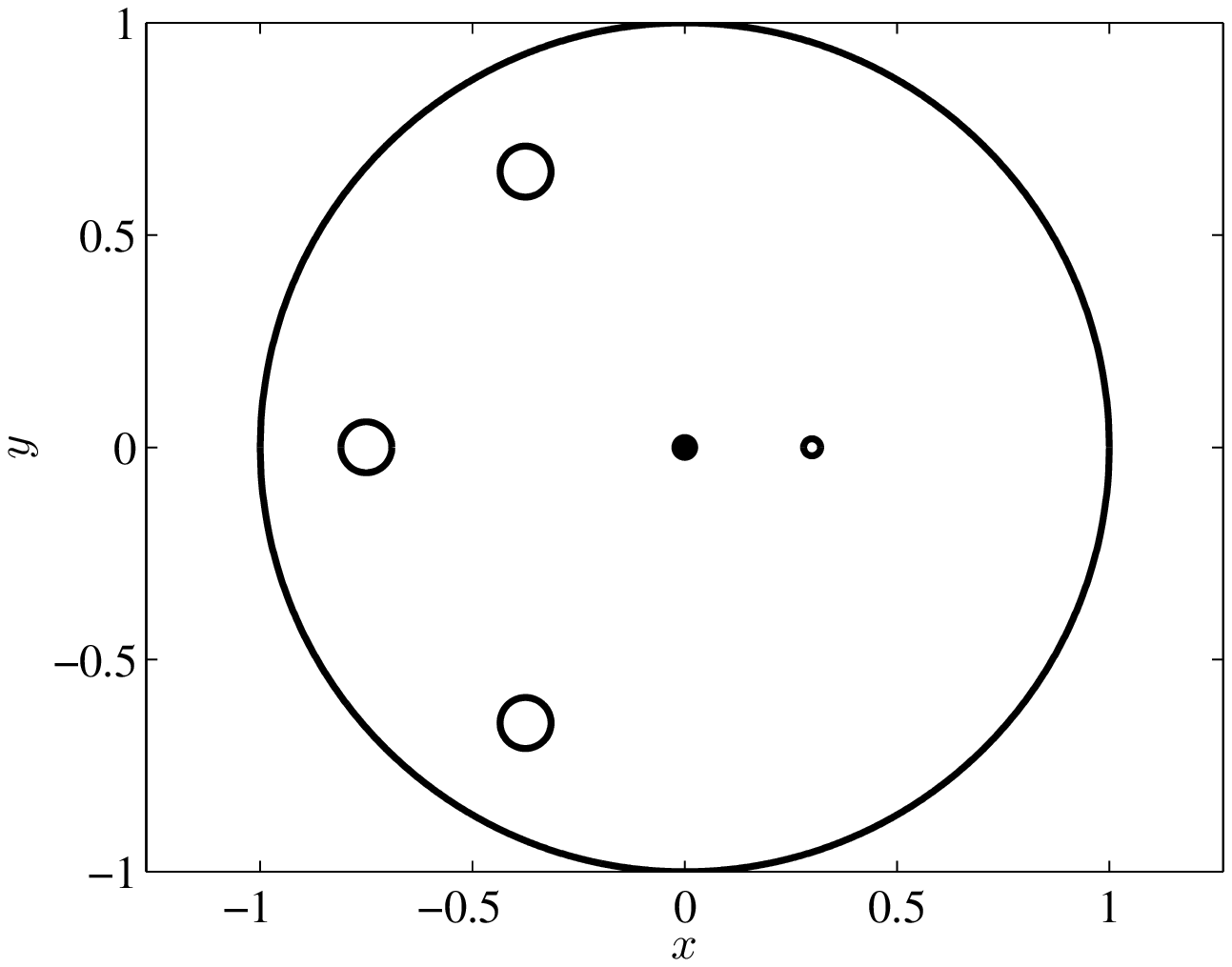}}
    \hspace{0.5cm}  
    \subfigure[Bimodal passage time density.]{\label{multimodediffsize}\includegraphics[width=.475\textwidth]{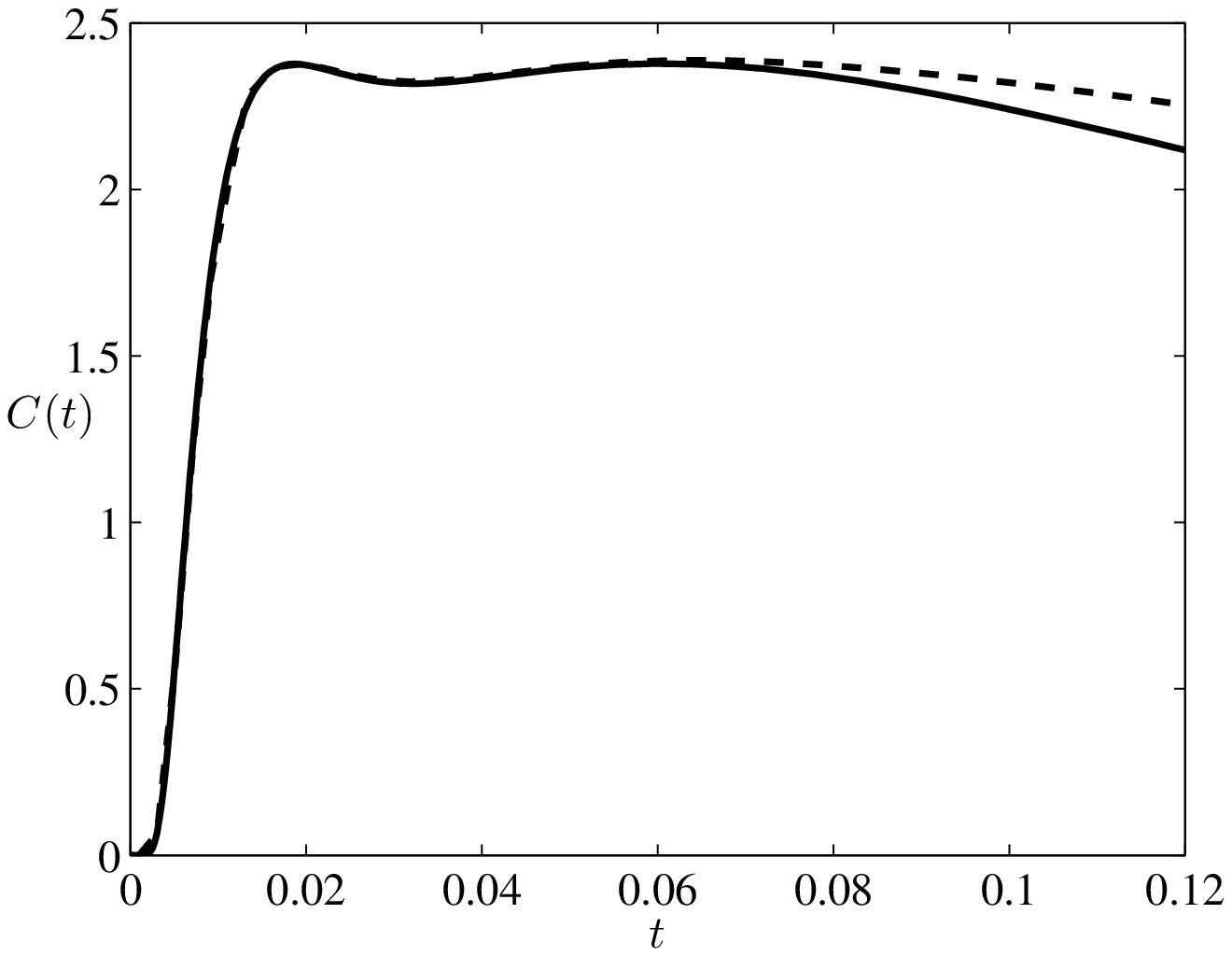}}
    \parbox{0.75\textwidth}{\caption{Schematic and results for Example \ref{multimode2}. In (a), we show the starting location $\bx_0$ at the origin (solid dot) surrounded by two absorbing sets. The first set consists of the nearest trap centered at $(r_1,0)$. The second set is composed of the union of the three traps centered on a ring $r_c > r_1$. In (b), we show the resulting bimodal distribution for $C(t)$. As in Example \ref{multimode1}, the ``wider net'' cast by the second absorbing set accommodates the increased dispersal of paths by the time they have reached the ring $r = r_c$. \label{fig:multimode2}}}
\end{figure}
In the final example, we demonstrate the efficacy of the two-term estimate for the case in which $\Omega$ is an ellipse. While the leading order term is independent of curvature, the second order analysis leading to \eqref{a1b1} must altered to account for the non-constant curvature, yielding in place of \eqref{a1b1},
\BE \label{a1b1gen}
	a_1 = -\phi\kappa + 2|\br|_\xi \frac{c_0^\prime}{c_0} - 2|\br|_\xi\frac{\phi^\prime}{\phi} + |\br|_{\xi\xi} \,, \qquad b_1 = -2\kappa|\br|_\xi^2-2|\br|_\xi\phi^\prime \,.
\EE
\noindent For an ellipse parameterized by 
\BE \label{paramellipse}
x = r_A\cos t \,, \qquad y = r_B\sin t \,; \qquad t \in \lbrack 0, 2\pi) \,,
\EE
\noindent with $r_A > r_B$, the curvature $\kappa(\xi)$ is defined implicitly in terms of the arc length $\xi$ by
\BE \label{curveellipse}
	\xi(t) = r_A \int_0^t \! \sqrt{1-k^2\cos^2 t} \, dt \,, \qquad \kappa(t) = \frac{r_A r_B}{\left(r_B^2\cos^2 t + r_A^2\sin^2 t \right)^{3/2}} \,; \qquad k \equiv \sqrt{1 - \left(\frac{r_B}{r_A} \right)^2} \,; \qquad r_A > r_B \,.
\EE
\noindent The rest of the analysis remains unchanged.

\begin{exmp} \label{ellipse}
In this example, we let $\Omega$ be an ellipse parameterized by \eqref{paramellipse} with $r_A = 1$ and $r_B = 1/2$. We place one circular trap of radius $\eps = 0.01$ centered at $\bx_1 = (0, 0.25)$ while setting the starting location at $\bx_0 = (0.7, 0)$. The schematic is shown in Fig.~\ref{ellipseschem}.
\end{exmp}

\noI In Fig.~\ref{ellipse3}, the numerical result for the capture time density $C(t)$ is shown in heavy solid, while the two-term estimate, using the modified coefficients \eqref{a1b1gen}, is shown in heavy dashed. To illustrate the boundary effect of the ellipse, we show in light solid the two-term estimate for the same trap and starting locations but with $\Omega$ replaced by the unit disk (light solid in Fig.~\ref{ellipseschem}). The difference clearly suggests that the portion of the boundary closest to the trap helps funnel particles toward the trap, allowing a greater portion of particles to the captured in early time. This effect is absent in the case of the unit circle on which the trap lies too far from the boundary. The funneling effect of the elliptical boundary is well-captured by the two-term estimate for the ellipse.

\begin{figure}[htbp]
\centering
    \subfigure[Example \ref{ellipse} schematic]{\label{ellipseschem} \includegraphics[width=.475\textwidth]{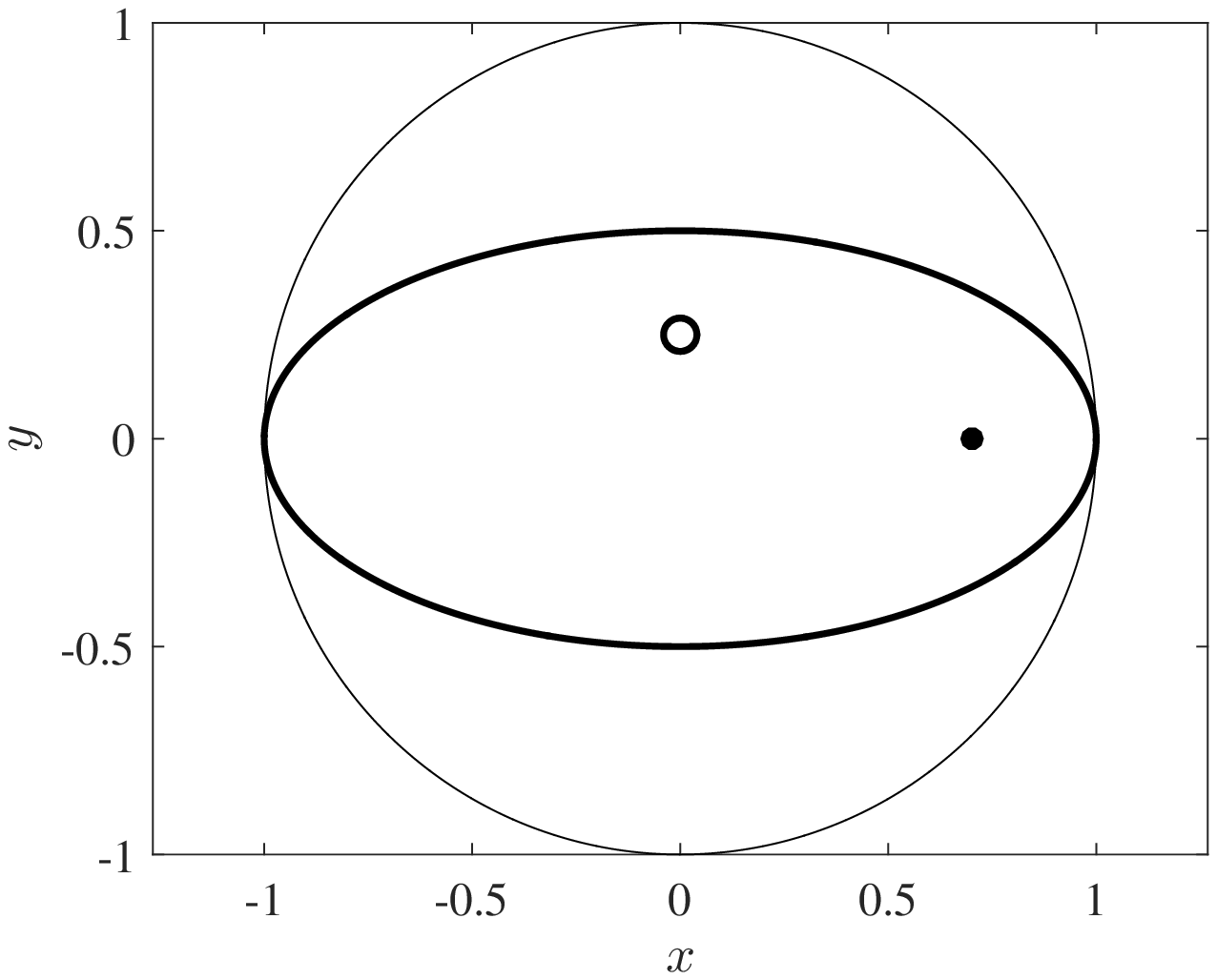}}
    \hspace{0.5cm}  
    \subfigure[$C(t)$]{\label{ellipse3}\includegraphics[width=.475\textwidth]{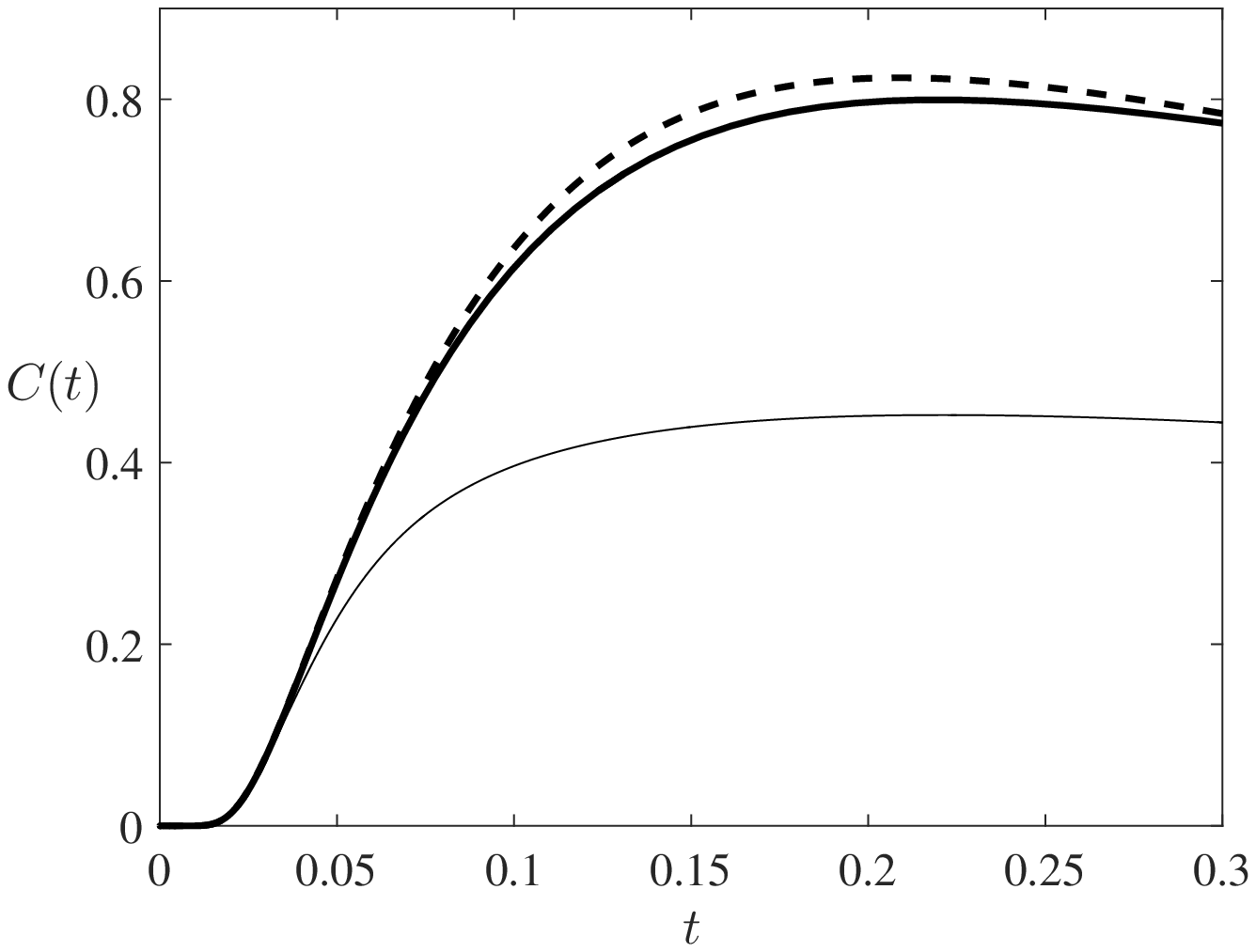}}
    \parbox{0.75\textwidth}{\caption{Schematic and results for Example \ref{ellipse}. In (a), we show the starting location $\bx_0 = (0.7, 0)$ (solid dot) and a circular trap of radius $\eps = 0.01$ centered at $\bx_1 = (0, 1/4)$. The elliptical (circular) domain is outlined in heavy (light) solid. In (b), we show the resulting distribution for $C(t)$, with the numerical result shown in heavy solid and the two-term estimate for the ellipse shown in heavy dashed. In light solid is the two-term estimate for the same trap and starting locations, but with the ellipse replaced by the unit disk. \label{fig:ellipse}}}
\end{figure}

In these examples, we have demonstrated the value of the two-term estimate in not only capturing boundary effects, but in predicting $P(t)$ and $C(t)$ to almost $t \sim \mathcal{O}(1)$ in some cases. We have also shown that it is able to capture the mode of the full capture time distribution, and, in cases where there are two distinct absorbing sets, the bimodal behavior of $C(t)$ in the early time evolution. We have also demonstrated its efficacy on a non radially symmetric domain.
%
%

\setcounter{equation}{0}
\section{Estimating passage time density at large times} \label{larget}

In the previous sections, we have focused on determination of the passage time density at short times where the starting location and geometry strongly influence the capture rate. At larger times, the free probability decays exponentially according to \eqref{LargeTimeApprox}, where $\lambda_0(\eps)$ is the principal eigenvalue of the Laplacian \eqref{eigproblem}. The quantity $\lambda_0(\eps)$ and its associated eigenfunction has been calculated (cf.~\cite{F,KTW,O,WHK,WK,Pillay2010,Ward2010,LTK2015}) as $\eps\to0$ and can be determined numerically by boundary integral equation methods (cf.~\cite{tai1974helmholtz}).
\begin{figure}[htbp]
\centering
\subfigure[Example 3.1]{\includegraphics[width = 0.47\textwidth]{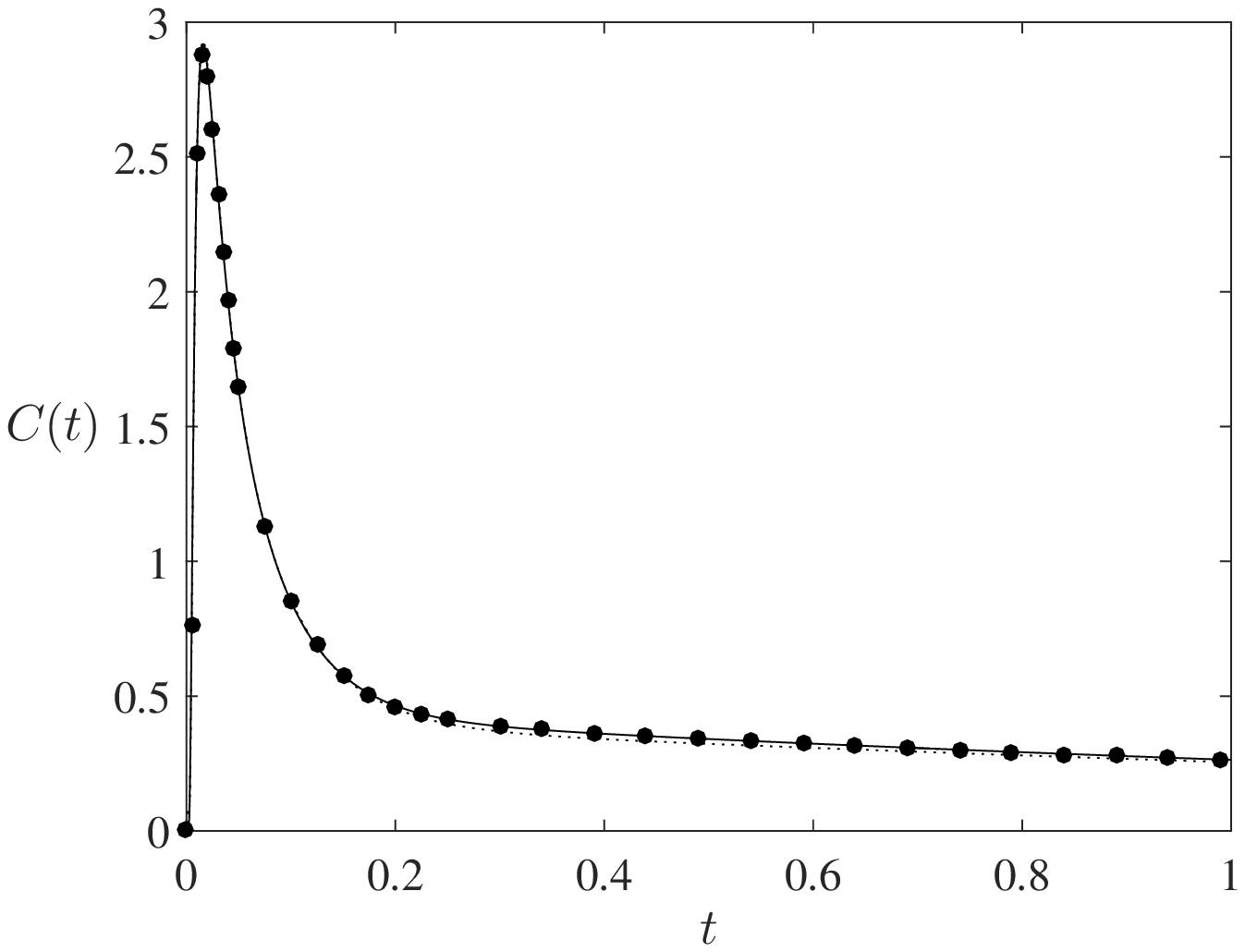}}\qquad
\subfigure[Example 4.3]{\includegraphics[width = 0.47\textwidth]{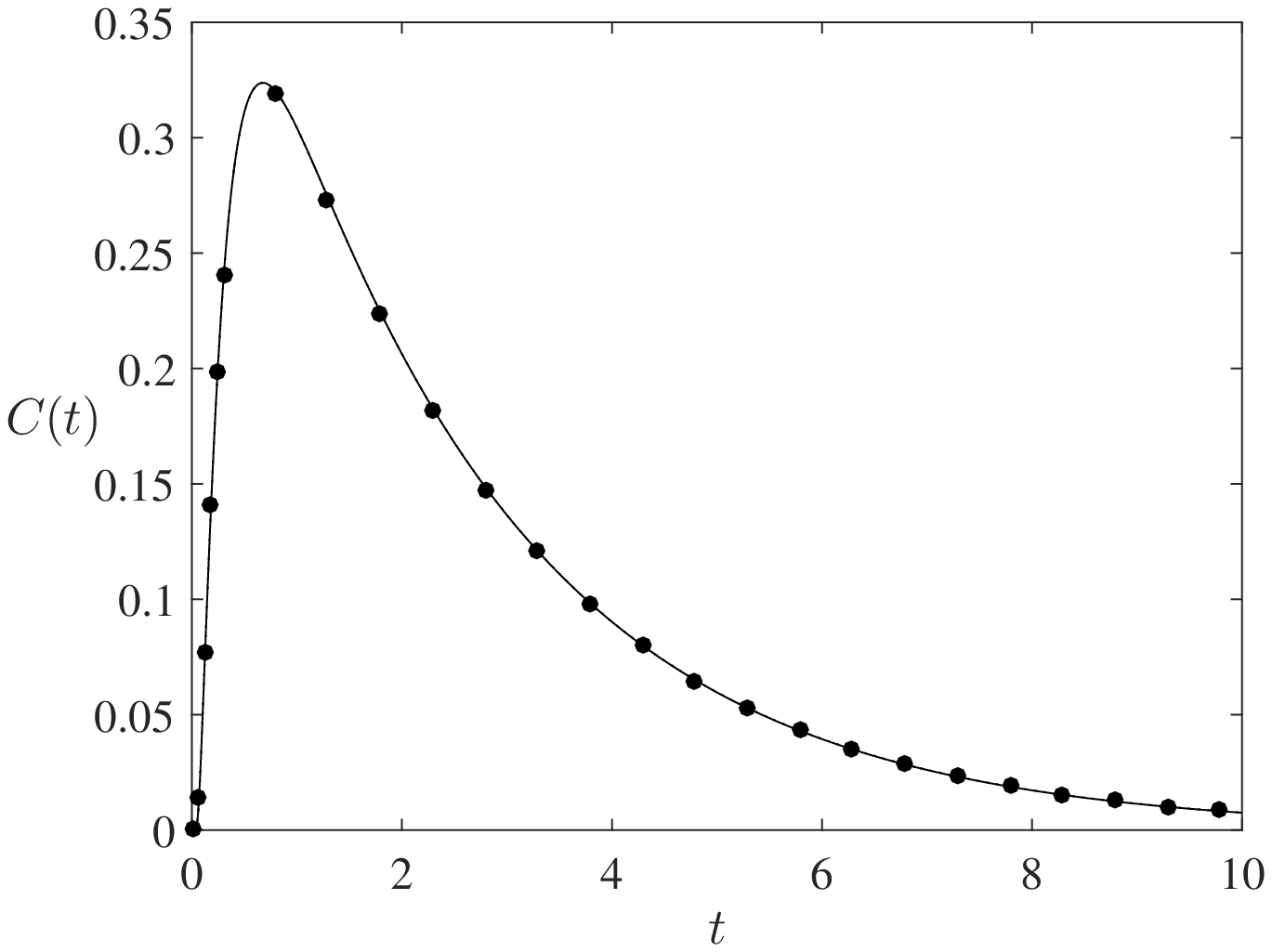}}\\
\subfigure[Example 4.6]{\includegraphics[width = 0.47\textwidth]{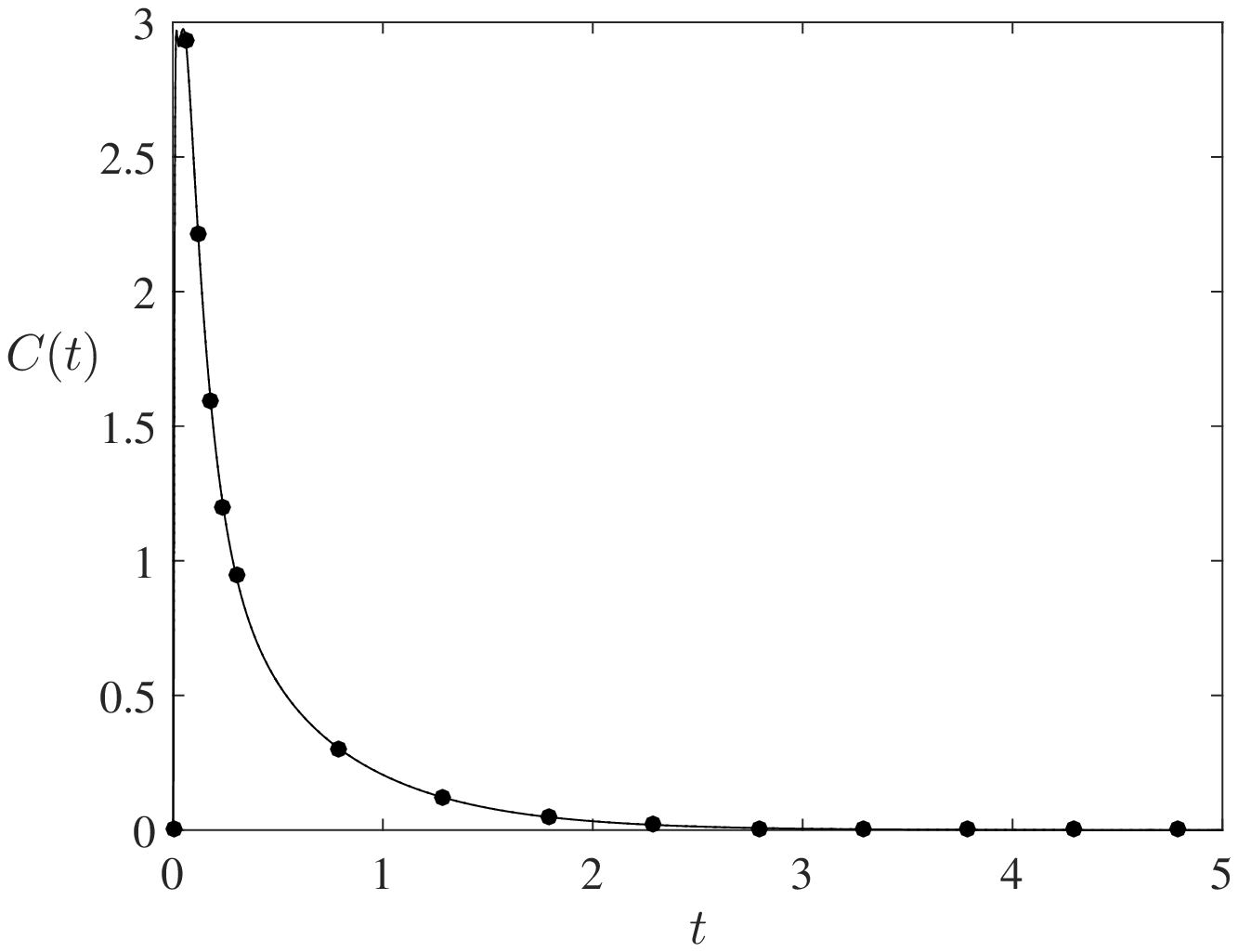}}\qquad
\subfigure[Example 4.7]{\includegraphics[width = 0.47\textwidth]{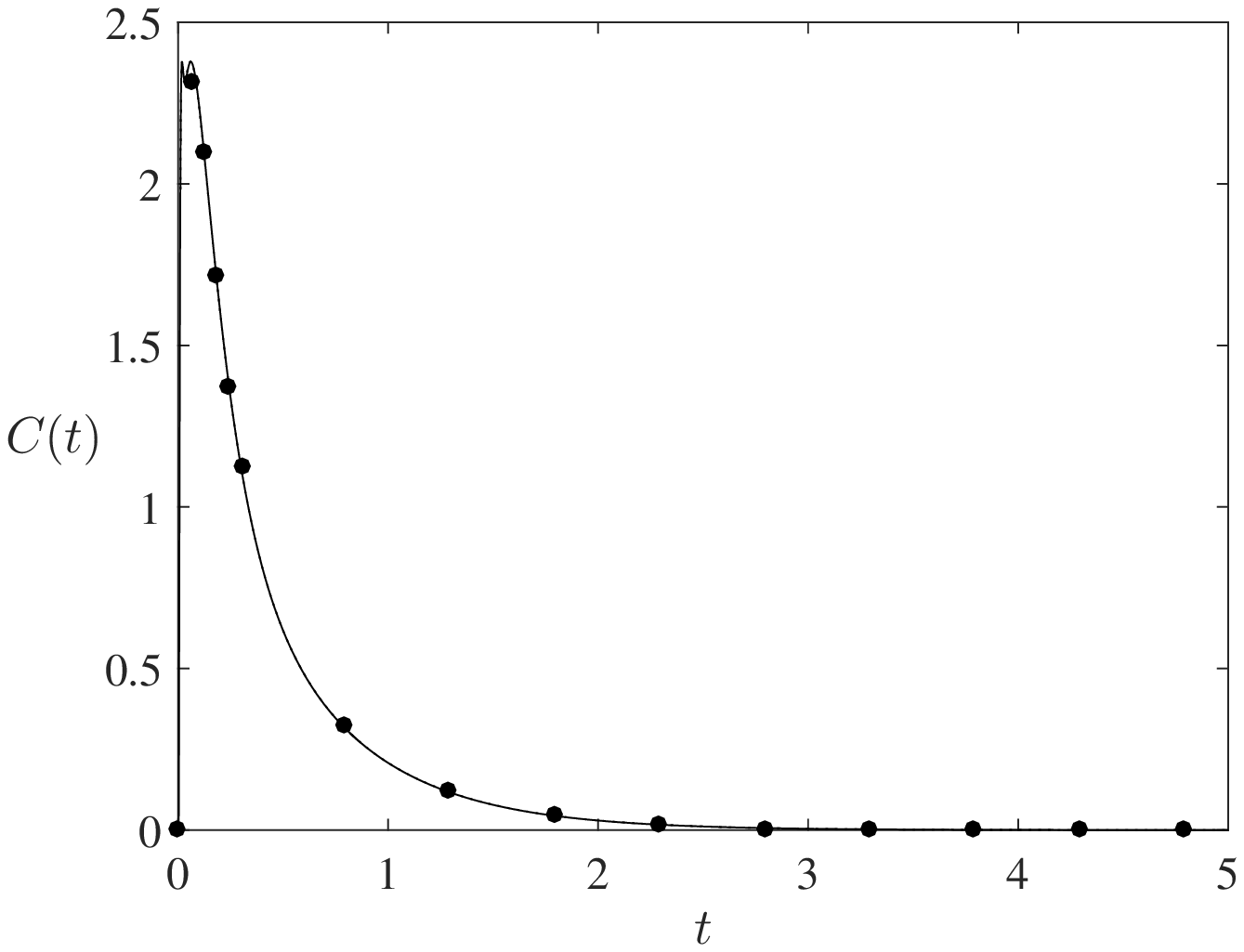}}
\parbox{0.75\textwidth}{\caption{The full distribution of passage times for the configurations of Examples 3.1, 4.3, 4.6 and 4.7 over large times. Comparison given for full finite element simulations (solid line) of \eqref{MainFreeProb}, the hybrid-asymptotic method (dashed line) and discrete particle simulations (solid dots). At this scale, the hybrid result is very difficult to distinguish from the exact numerical result. \label{fig:FullDist}}}
\end{figure}

Here we demonstrate the ability of the hybrid-asymptotic method to obtain the full capture time distribution $C(t)$ over larger ranges of $t$. In contrast to \S \ref{smallt}, where $G_h(\bx;\bx_j;s)$ and its regular part $R_h$ were evaluated by an asymptotic analysis of \eqref{Req} valid only in the limit $s\to\infty$, we use here a numerical finite element evaluation of \eqref{Req} that is valid for the entire relevant range of $s$. In Fig.~\ref{fig:FullDist} we display the agreement between full finite element simulations of \eqref{MainFreeProb}, the hybrid-asymptotic method of \S\ref{sec:FullDist} and discrete particle simulations (cf.~Appendix \ref{A:particles}) with very good agreement between the three methods.

\setcounter{equation}{0}
\section{Discussion} \label{discuss}

We have presented and demonstrated a hybrid asymptotic-numerical method for estimating the full capture time distribution $C(t)$ of the two-dimensional narrow capture problem with internal traps. The motivation for this work is a calculation of the variance of the MFPT (cf. \S\ref{sec:variance}) which is found to be asymptotically equal to the MFPT. This implies that the MFPT is not necessarily a reliable estimate of typical capture times therefore requiring a method for obtaining the full distribution. The method developed relies on accurate determination of a Helmholtz Green's function $G_h$ and in particular its regular part $R_h$. For $t \ll 1$, we calculate a two-term boundary layer asymptotic solution for $R_h$ and show that it can predict $C(t)$ for over moderately small times. When compared to a boundary-free approximation, the two-term estimate clearly shows the effect of the boundary funneling particles toward the trap(s), and also the timescale over which this effect becomes dominant. In addition, it is able to capture the bimodal nature of $C(t)$ in cases when the traps are arranged into two distinct absorbing sets. Finally, we have shown that the method also works on domains that do not exhibit radial symmetry. We remark that in this small $t$ regime, modulo the numerical inversion of the Laplace transform, our method for computing $C(t)$ requires only the asymptotic expansion of a certain Green's function.

For $t \sim \bigoh(1)$, the hybrid method can be employed by using numerical evaluation of $G_h$ and $R_h$. We remark that the method is accurate enough to capture some of the critical small time properties of $C(t)$, including when $C(t)$ is bimodal. We emphasize that we are able to accurately estimate statistics of a time-dependent stochastic processes in terms of a single time-independent Green's function. Furthermore, the estimates that we obtain are accurate beyond all orders of $\nu$, where $\nu \sim \mathcal{O}(-1/\log\eps)$ and $\eps$ is the typical trap radius.

An interesting problem would be to extend the techniques presented here to a three-dimensional geometry with multiple absorbing traps \cite{IN2013}. The boundary layer solution in the $t \ll 1$ regime would be a far more involved calculation. For $t \sim \bigoh(1)$, the Green's function solver would need to be not only accurate enough to resolve the $\bigoh(\eps)$ scale of the traps, but fast enough to be called repeatedly by the numerical Laplace inversion routine. 

Another open problem would be to compute the capture time density in cases in which the traps are non-static in time. Examples include scenarios in which the traps are mobile (e.g., \cite{tzou2015mean, tzou2014first}), or undergo stochastic switching between absorbing and non-absorbing states (\cite{bressloff2015stochastically}). In Fig.~\ref{onetrapschem_rotate}, we show a simple motion in which one small trap rotates concentric with the unit disk while the starting location of the particles $\bx_0$ is set close by. The resulting capture time density features multiple modes corresponding to the capture of particles on each successive sweep of the trap through $\bx_0$. As the particles disperse, the peaks in $C(t)$ become less localized in time. Multimodal distributions are also expected to arise in the aforementioned case of one stochastically switching trap, with the peaks occurring at times during which the trap is in the absorbing state. In such scenarios with multiple modes, the MFPT, or global MFPT, is even less informative than in the case of static traps. As such, it would be useful to develop techniques for obtaining full capture time distributions in cases when traps do not remain static in time. 

\begin{figure}[htbp]
\centering
    \subfigure[schematic for rotating trap]{\label{onetrapschem_rotate} \includegraphics[width=.475\textwidth]{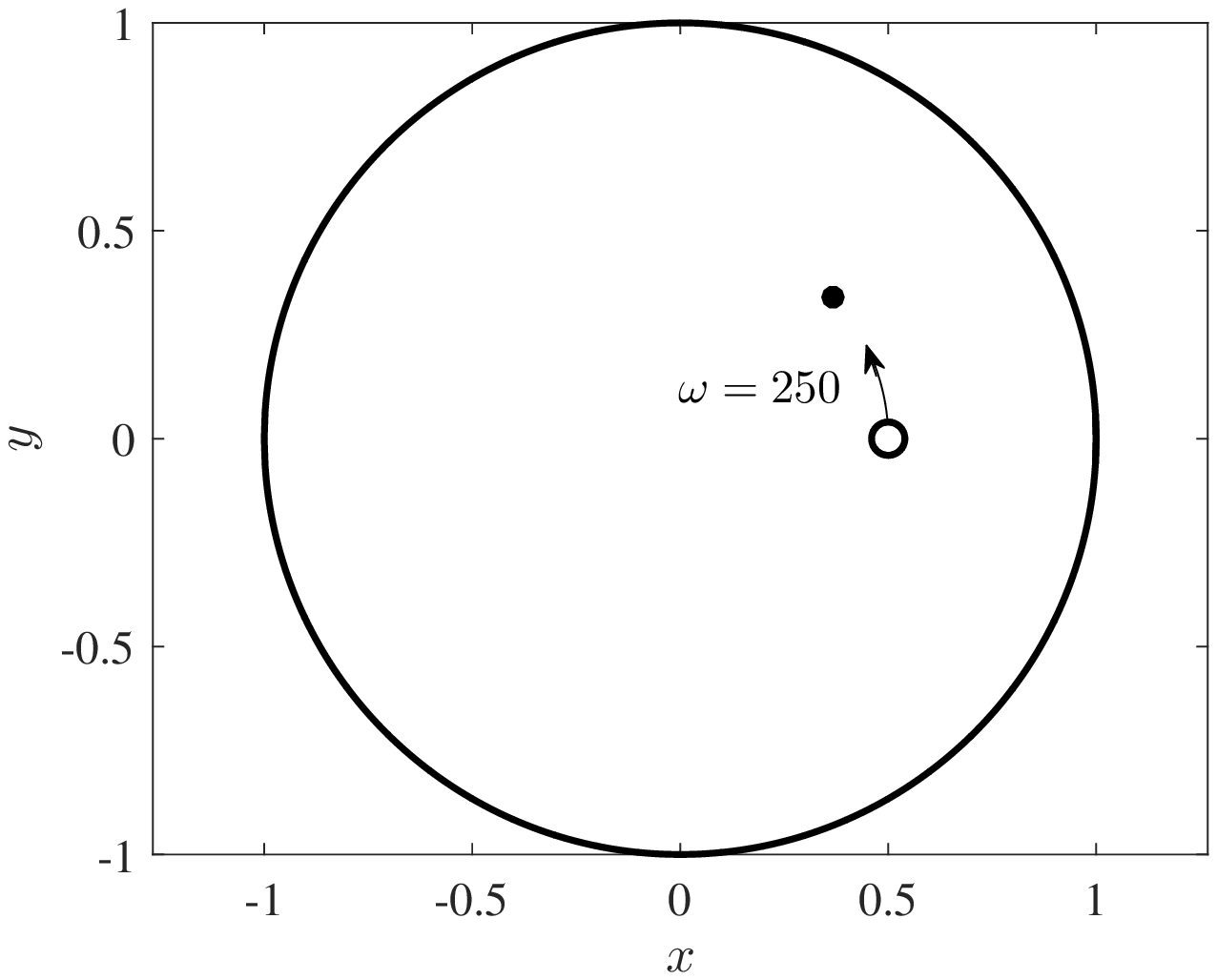}}  
    \hspace{0.5cm} 
    \subfigure[$C(t)$]{\label{mobile_C}\includegraphics[width=.475\textwidth]{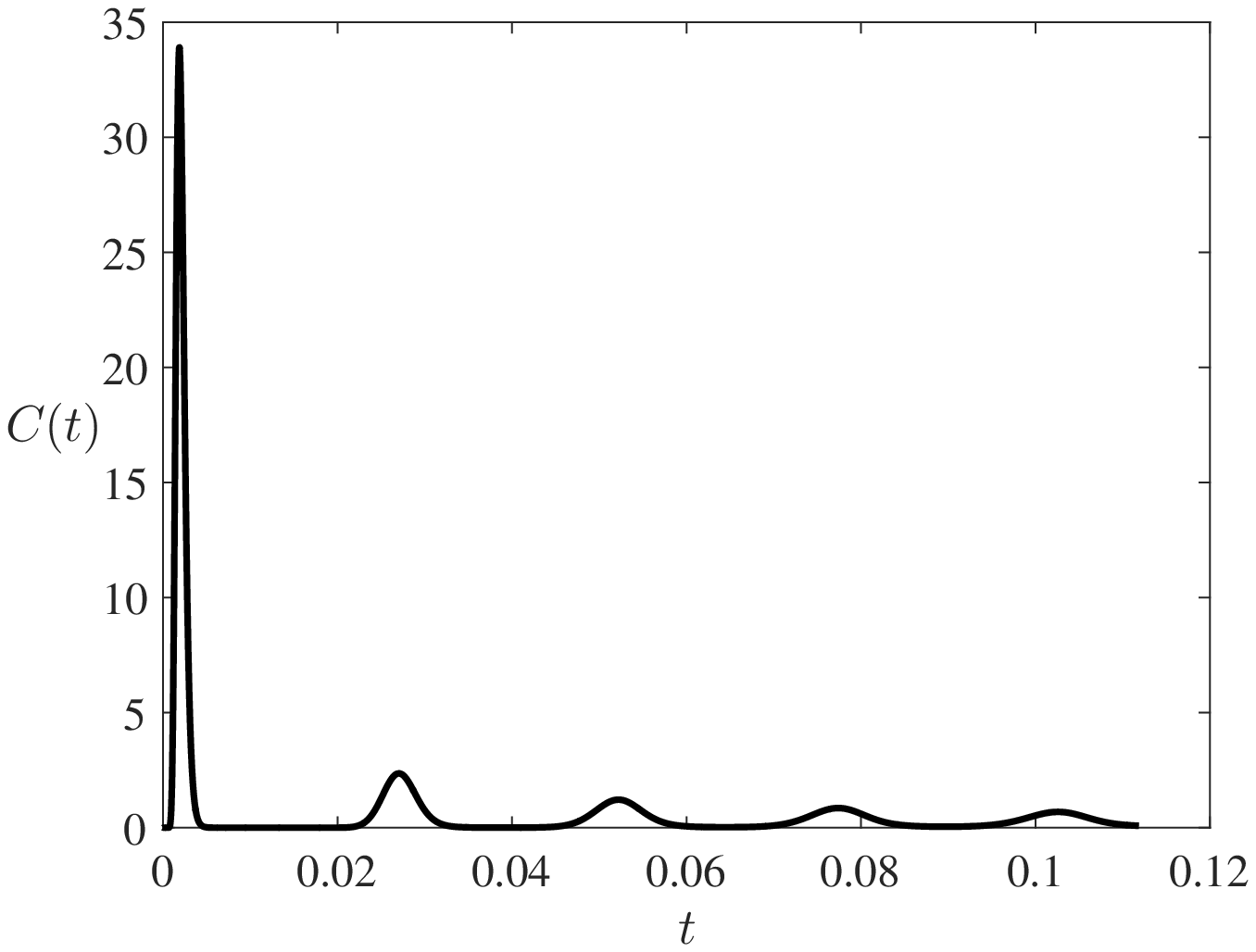}}
    \parbox{0.75\textwidth}{\caption{In (a), we show a schematic of a counterclockwise rotating trap of radius $\eps = 1\times 10^{-5}$ starting at $\bx_0 = (0.5, 0)$, while the starting location of the particles is at $(0.5\cos(0.5), 0.5\sin(0.5))$. In (b), we show the capture time density $C(t)$. The peaks correspond to the capture of particles on each successive sweep of the trap through $\bx_0$.  \label{fig:mobile}}}
\end{figure}

\section*{Acknowledgments}
A.E.L. and R.T.S. acknowledge support from NSF grant DMS-1516753. J.C.T. was partially supported by a PIMS CRG Postdoctoral Fellowship. The authors gratefully acknowledge the insightful contributions of Andrew Bernoff, Theodore Kolokolnikov, and Michael J. Ward. A.E.L acknowledges the assistance of the Notre Dame Center for Research Computing (CRC).

\begin{appendix}

\section{Particle Simulations}\label{A:particles}

The particle simulations of Fig.~\ref{fig:FullDist} involved 2 million discrete Brownian paths. When particles encountered the outer boundary, they were reflected back into the domain at an angle equal to their incidence angle with respect to the normal vector at the boundary contact point. An adaptive time step was used based on shortest distance to an absorbing set which allows for high accuracy close to absorption while accelerating the sampling of long excursions. Simulations were run in parallel using facilities at the Notre Dame Center for Research Computing (CRC).

\end{appendix}

\bibliographystyle{elsart}
\bibliography{narrowbib}

\begin{thebibliography}{10}
\bibitem{supp}
{S}ee {S}upplemental {M}aterial at (url will be inserted by publisher) for
  {MATLAB} finite elements code that reproduces results of selected examples.
\end{thebibliography}

\end{document}